\pdfoutput=1
\documentclass[11pt]{article}
\usepackage[T1]{fontenc}
\usepackage[utf8]{inputenc}
\usepackage{lmodern}
\usepackage{microtype}
\usepackage{amsmath,amssymb}
\usepackage{graphicx}
\usepackage{placeins}
\usepackage{booktabs,array}
\usepackage[a4paper,margin=2.5cm]{geometry}
\usepackage[numbers,sort&compress]{natbib}
\usepackage[hidelinks]{hyperref}

\title{Identification and Electrothermal Modeling of Persistent Current Switch Operating Regimes in a Siemens MRI Magnet Using External Electrical Measurements}
\author{Volodymyr Pupkov\thanks{Disclosure: the author is an employee of Digit Systems, the manufacturer of the MPS2 service power supply that recorded the analyzed data. The analysis was performed by the author; Digit Systems approved the publication of derived results. Siemens, the magnet manufacturer, was not involved in this work. No customer-identifying information is used or disclosed.}\\[2pt]\normalsize Digit Systems\\\normalsize ORCID 0009-0004-5061-5625}
\date{}
\hypersetup{pdftitle={Identification and Electrothermal Modeling of Persistent Current Switch Operating Regimes in a Siemens MRI Magnet Using External Electrical Measurements}, pdfauthor={Volodymyr Pupkov}, pdfkeywords={persistent-current switch; superconducting MRI magnet; electrothermal modeling; self-heating; retrapping; magnet ramping}}

\begin{document}
\maketitle

\begin{abstract}
The persistent current switch (PCS) closes the superconducting loop of an MRI magnet. During ramping it is commonly treated as a binary element: with the heater on, the PCS branch is assumed effectively open; after heater turn-off, it is assumed superconducting. Field measurements with an MPS2 service power supply (Digit Systems) on a Siemens MAGNETOM Avanto show that the PCS remains electrically active and that its resistive state can persist after heater turn-off through Joule self-heating.

Dedicated maneuvers identified two resistive regimes. With the heater on, the low-voltage resistance is nearly constant; after heater turn-off, self-heating produces a power-law voltage dependence. Both are described by a single effective electrothermal relation, $R_{sw}=C\,(P_h h+V_{node}^2/R_{sw})^a$, where $h$ is the recorded heater state, spanning a 690-fold resistance range.

Final-hold data from 17 Siemens magnet sessions confirmed the voltage dependence of effective PCS resistance. Since branch-current relaxation follows $\tau=L/R_{sw}$, the required hold time depends on switch state rather than a fixed delay; the median margin after source settling was $3.6\,\tau$.

The intersection of the PCS characteristic with the external-circuit load line defines the boundary of self-sustained resistive operation. For Avanto, the predicted 0.062 V threshold lay between superconducting recovery at 0.06 V and resistive-state persistence at 0.07 V, and retrospectively classified all 24 archived segments correctly.

Including both PCS resistive regimes in the dynamic model reproduced the observed current redistribution: RMSE decreased from 1.449 to 0.154 A, a factor of 9.4. This improvement required explicit representation of heater state.
\end{abstract}

\tableofcontents
\bigskip

\FloatBarrier
\section{Introduction}
In persistent mode, the current of an MRI magnet circulates in a closed superconducting circuit formed by the coil and the persistent current switch (PCS) connected in parallel with it. To change the current, the heater drives the PCS into a resistive state and connects the coil to an external power supply. The PCS itself is located inside the cryostat: its temperature and branch current are not measured directly, while the service power supply records only the total current and external voltages. In the simplest model, the heated PCS is treated as an open circuit and the power-supply current is identified with the coil current.

A violation of this approximation first became apparent in additional test maneuvers performed after a service operation to clarify the magnet behavior and subsequently improve ramping profiles (Fig. 1).

\begin{figure}[!htbp]
\centering
\includegraphics[width=\linewidth]{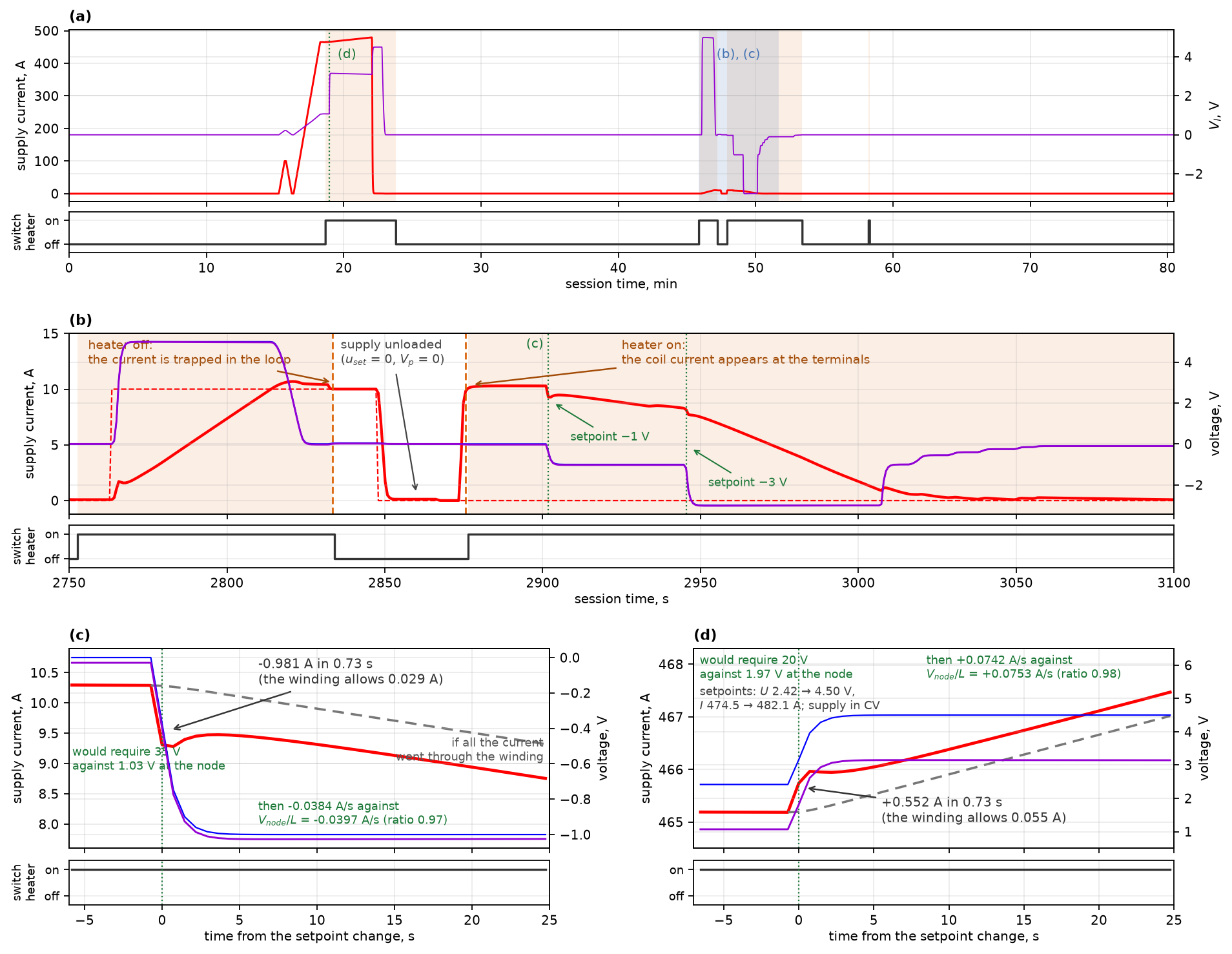}
\caption{Field observation of the difference between power-supply current and coil current on a Siemens MAGNETOM Avanto 1.5 T. Red solid curve: power-supply current; red dashed curve: its setpoint; blue and purple curves: power-supply and magnet voltages; the lower trace shows the PCS heater state. (a) Full session: a repark interrupted by a current collapse near 480 A with the heater on (the cause is not established here; such collapses are excluded from the Section 6.3 corpus), followed by a low-current diagnostic maneuver. (b) Maneuver at 10 A: current capture into the persistent loop, unloading of the power supply, and subsequent opening of the PCS by turning the heater on. (c) After a voltage step at 10 A, the power-supply current changes much faster than permitted by the coil inductance; the subsequent rate is consistent with $V/L$. (d) A similar fast component is observed during a routine repark near 465 A, after which the dynamics again follow $V/L$.}
\label{fig:1}
\end{figure}
In one of these maneuvers, the coil current of a MAGNETOM Avanto 1.5 T was brought to approximately 10 A, after which the PCS heater was switched off to trap the current in the superconducting loop ($V_{node}\approx 0$); the power-supply current was then reduced to zero, and the heater was switched on again to bring the circulating current back to the external terminals (Fig. 1b).

At this low current, the difference between power-supply current and coil current became especially clear. During a voltage step, the measured power-supply current changed by 0.981 A in 0.73 s, whereas the observed voltage was sufficient to change the coil current by only about 0.029 A over the same interval (Fig. 1c). After the fast transition, the subsequent current-change rate nearly coincided with $V/L$, i.e. with the expected inductive dynamics of the coil. This indicated a fast current component that did not belong to the coil.

The question was whether the same effect was present during ordinary operation at currents of hundreds of amperes, where a change of about one ampere is almost invisible on a full-scale plot. Magnification of a repark segment near 465 A showed the same behavior (Fig. 1d): after a voltage-setpoint change, the power-supply current increased by 0.552 A in one logging step, whereas the coil could have changed by only about 0.055 A. After this fast response, the subsequent increase was 0.0742 A/s, compared with the calculated $V/L=0.0753$ A/s. Thus, the same signature is observed both in the deliberately performed low-current maneuver and in a routine repark: rapid changes in power-supply current cannot be explained by coil inductance, whereas the subsequent slow component is consistent with it.

These observations require the PCS to be treated as a separate electrical branch. The present work therefore examines the characteristic $R_{sw}=R(V,h)$, where $h$ is the recorded heater state. For $h=1$, external heating establishes one resistive regime; after the heater is turned off, the already formed normal zone may persist because of Joule self-heating. Thus, $h$ is a control input but not a label of the PCS state: at $h=0$, the PCS may be either superconducting or resistive. Return to the superconducting state when self-heating is insufficient is referred to below as retrapping.

Determining the characteristic of the path that closes this circuit, and its dependence on heater state, from external electrical measurements is the subject of this work. The terminology used throughout the paper is therefore fixed here: $h$ reports only the heater state, and the time of each recorded heater-state change is known to within one CAN polling interval; the PCS state itself is a separate history-dependent variable, and its transitions need not coincide with a change in $h$. After heater switch-off, the resistive state was observed to persist for another 59 s (Section 5.1). Distinguishing these two quantities is central to the following sections.

Quantitative identification was performed on one PCS of a Siemens MAGNETOM Avanto. The work includes methods for externally separating PCS and coil currents, measurement of two resistive regimes, a criterion for self-sustain of the resistive state, and validation of a dynamic model; logs from other magnets are used to analyze general consequences of the circuit topology and the final relaxation.

\FloatBarrier
\section{State of the Art}
The study did not begin from a preselected electrothermal model of the PCS. The observed effects were first found in field data while constructing an engineering model of the power-supply-to-magnet path and during subsequent diagnostic maneuvers. After they had been quantitatively isolated, the results were compared with the literature on normal-zone dynamics in superconductors. This comparison provided a physical interpretation of the observations and at the same time defined the novelty boundary of the present work: the mechanisms of self-heating, bistability, and retrapping are known, whereas the subject here is their identification from external electrical measurements of an installed PCS.

Self-heating of a superconducting normal zone can produce a multivalued hysteretic current-voltage characteristic; stable operating points are determined by its intersection with a load line \cite{r1,r2}. As current decreases, return of the hot zone to the superconducting state is described as retrapping \cite{r3}, while persistence of a resistive state under electrothermal feedback is described as latching \cite{r4,r5}. In the present work, neither temperature nor the structure of the normal zone is measured; retrapping is therefore identified only from the external electrical transition.

Classical cryostability theory considers the thermal balance of a superconductor at a prescribed current \cite{r6}. Here, the PCS current itself depends on node voltage, branch characteristic, coil current, and the resistance of the external path. Therefore, the self-sustain boundary observable through external electrical variables is determined jointly by the PCS characteristic and the connected external circuit; it is not solely an internal PCS parameter.

Published studies report the resistance of thin-film PCS devices as a function of heater power \cite{r7}, heater requirements for REBCO switches \cite{r8}, and thermal characteristics of CuNi-NbTi MRI-class PCS devices \cite{r9,r10}. For the latter, two 90-$\Omega$ thermofoil heaters placed between layers of a bifilar winding and a total normal resistance of the PCS itself of 12.5 $\Omega$ at 15 K are described. These quantities refer to a different construction: 90 $\Omega$ is the electrical resistance of the heater, whereas 12.5 $\Omega$ is the resistance of the normal zone, and neither is directly comparable with the effective branch resistance of the Avanto PCS studied here. Addition of heater power and Joule dissipation is also established physics; the novelty of this work is not a new thermal mechanism, but its quantitative identification from external data recorded by the MPS2 service power supply.

Open sources describe the ramp sequence mainly qualitatively. No quantitative analysis of current separation, final relaxation, and the self-sustain boundary of an installed PCS using service logs was found in the accessible literature. This statement refers only to open publications; proprietary manufacturer service procedures may contain additional practical knowledge.

\FloatBarrier
\section{Experimental Object, Measurement Circuit, and Data}
\FloatBarrier
\subsection{Experimental object and scope of applicability}
Quantitative identification was performed on the PCS of a Siemens MAGNETOM Avanto magnet accessible for bench service maneuvers with a current limit of 50 A, compared with an operating value of approximately 500 A; the actual current in these maneuvers did not exceed 42 A. This limit applies specifically to the identification maneuvers: archived records from other magnets of the same family include routine operation at full current, but coefficients are not identified from those data. The numerical resistance laws and electrothermal-balance parameters were determined under these conditions; archived data from other magnets are used to test observable consequences of the model.

\FloatBarrier
\subsection{Electrical circuit and measurement access}
The PCS is connected in parallel with the coil; the MPS2 service power supply (Digit Systems), referred to below as the \emph{power supply}, is connected to the external current terminals of the magnet. Positive $I_{psu}$ is defined from the power supply toward the magnet, and voltage polarities are chosen so that positive $V_{node}$ increases the coil current. The currents and voltages satisfy

\[
I_{psu} = I_{coil} + I_{sw},
\qquad
L\frac{dI_{coil}}{dt}=V_{node},
\qquad
I_{sw}=\frac{V_{node}}{R_{sw}},
\]
\[
V_{node}=V_l-R_{lead}I_{psu}
=V_p-R_{path}I_{psu},
\qquad
R_{path}=R_{cab}+R_{lead}.
\]
Here, $V_p$ is the measured voltage at the power-supply terminals (the commanded voltage setpoint is denoted below by $V_{set}$ and is not identical to $V_p$); $V_l$ is the voltage measured at the external magnet terminals, at the end of the connection line from the power supply; $R_{cab}$ is the resistance of the power-supply cable; $R_{lead}$ is the resistance of the current leads from the magnet terminals to the internal node; and $R_{path}$ is their sum. $R_{sw}$ denotes the effective resistance of the resistive PCS branch, not the electrical resistance of its heater. The node and measurement-access points are shown in Fig. 2.

\begin{figure}[!htbp]
\centering
\includegraphics[width=\linewidth]{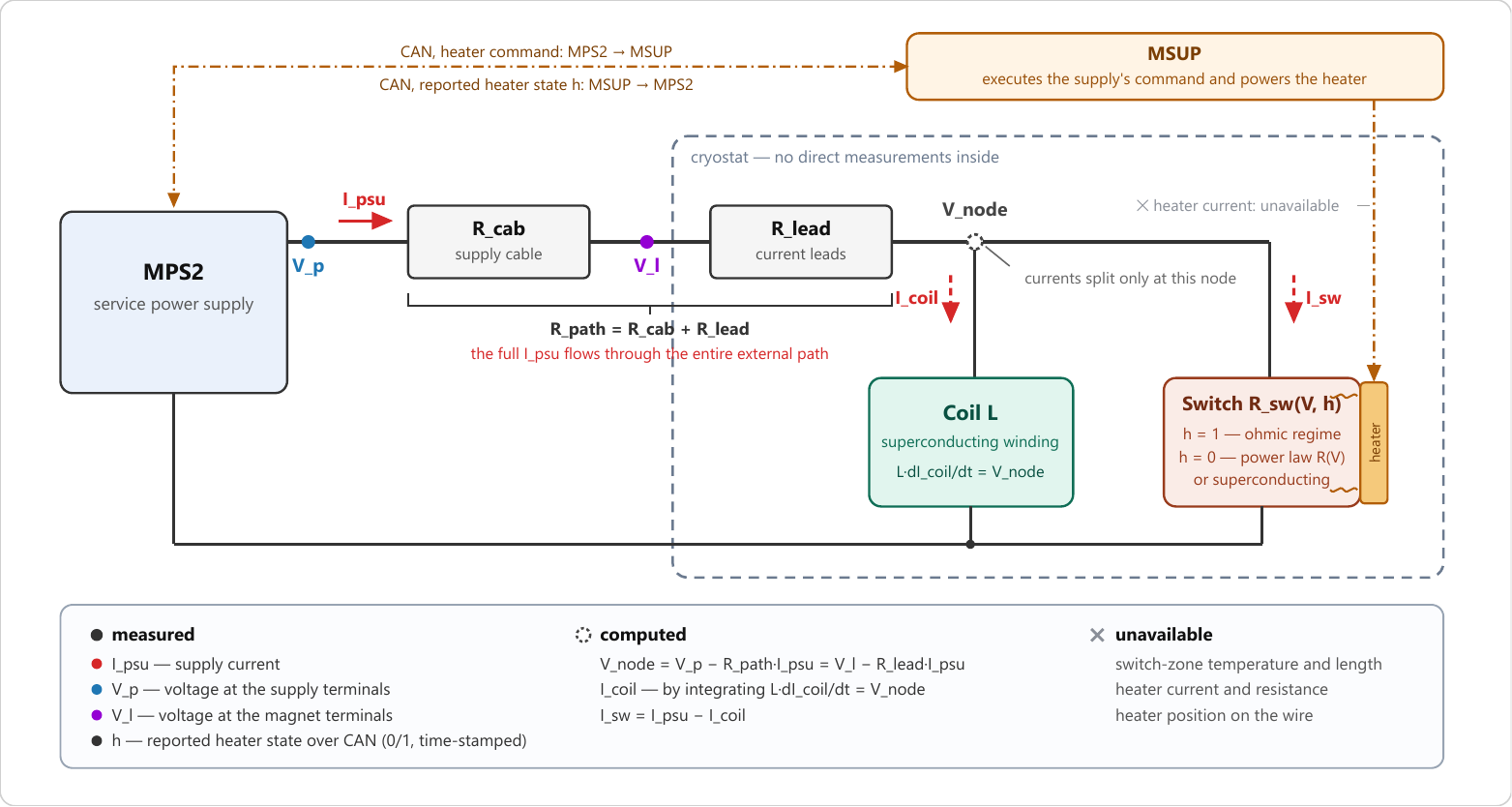}
\caption{The coil $\parallel$ PCS node and measurement access. The recorded quantities are $I_{psu}$, $V_p$, $V_l$, and heater state $h$; $V_{node}$ and the branch currents are calculated using $R_{path}=R_{cab}+R_{lead}$. The full $I_{psu}$ flows through the entire external path: separation into $I_{coil}$ and $I_{sw}$ occurs only at the internal node. The gray cryostat region is not directly accessible to measurement.}
\label{fig:2}
\end{figure}
\FloatBarrier
The PCS heater is controlled by the Magnet Supervision module (MSUP), the standard Siemens module for monitoring and controlling magnet functions. The power supply communicates with it over CAN: it sends a command to turn the heater on or off, the MSUP executes the command, powers the heater, and manages the process, while a separate request from the power supply polls the heater state and records the reply. This recorded value $h\in\{0,1\}$ is used below as a control input of the model. Heater current, electrical resistance, and temperature are not observed directly; the available data cannot distinguish whether the reported state represents the actual power dissipation in the heater or only the command accepted by the MSUP (Section 8). The work characterizes the action of the heater on the PCS, not the heater construction itself.

The Avanto parameters used below were determined as follows:

\begin{center}
\begin{tabular}{lrl}
\toprule
Quantity & Value & Method \\
\midrule
Inductance $L$ & 26.39 H (0.4\% range) & five independent cycles without a PCS model (Section 4.2) \\
$R_{lead}$ & 2.56--2.59 m$\Omega$ & decay tail in current-limit mode \\
$R_{cab}$ & 2.24 m$\Omega$ & cable test on a shorted loop \\
\bottomrule
\end{tabular}
\end{center}
For the calculations below, $R_{path}=4.802$ m$\Omega$ is used. This value is consistent with the sum $R_{cab}+R_{lead}$ and is independently confirmed by three episodes with $V_{node}\approx 0$, agreeing within 0.001 m$\Omega$.

Two conventions are essential when reconstructing $V_{node}$. First, the full current $I_{psu}$, not only the coil current, passes through $R_{cab}$ and $R_{lead}$. The reason is topological: current divides into $I_{coil}$ and $I_{sw}$ only at the internal node, downstream of the current leads, so the entire external path --- cable and current leads --- carries their sum, the full $I_{psu}$; the only branching point lies inside the cryostat (Fig. 2). Using $I_{coil}$ in the resistive correction distorts the characteristic: the relative error associated with the omitted term $R_{lead}I_{sw}$ is about 0.5\% at $V_{node}=0.8$ V and 21\% at 0.07 V. The consequences of the alternative reconstruction are discussed in Section 7.5.

Second, the subtraction is signed: $V_{node}$ is not the ``power-supply voltage minus losses in magnitude,'' but the algebraic residual after the external-path drop, and its sign has physical meaning. During ramp-up, the external path reduces the node voltage available to the magnet, while during ramp-down, when voltage and current have opposite signs, it increases it. Replacing signed quantities by magnitudes reverses the effect and changes the system position relative to the retrapping boundary. The usual reconstruction of $I_{coil}(t)$ is obtained by integrating $V_{node}/L$ and requires an initial condition. The initial condition is set from zero-setpoint windows with $V_{node}\approx0$. This does not contradict Section 4: specially designed methods first determine the target quantities independently of any assumed initial current and are then used to validate the integral reconstruction.

\FloatBarrier
\subsection{Dedicated Avanto sessions}
All dedicated maneuvers were performed on one Avanto PCS. In all rows below, $V_{set}$ denotes the voltage setpoint of the power supply; the measured power-supply terminal voltage is denoted by $V_p$ (Section 3.2) and differs from the setpoint.

The $V_{set}$ levels in these maneuvers were chosen based on measurability requirements and fixed before the maneuvers were performed. Determination of steady-state characteristics requires long plateaus, while the coil integrates the node voltage throughout. Therefore, using the normal operating voltage would have driven the current to the allowable limit long before the protocol was complete. With $L=26.39$ H, a setpoint of about 10 V gives $dI_{coil}/dt\approx0.38$ A/s; the measured value on this magnet is 0.378--0.379 A/s. At that rate, the 50 A limit is reached in about 2.2 min, whereas the protocols used here last 14--88 min. The voltage was therefore deliberately reduced for the dedicated maneuvers. Lower voltage levels give sequences of roughly fifteen segments enough time to reach steady values within the allowed current range: the current limit imposes the voltage limit here; the two were not selected independently.

\begin{center}
\begin{tabular}{>{\raggedright\arraybackslash}p{3.9cm}>{\raggedright\arraybackslash}p{5.7cm}>{\raggedright\arraybackslash}p{4.2cm}}
\toprule
Maneuver & Quantity determined & Role in analysis \\
\midrule
sequence of $V_{set}$ levels: 1.0 $\to$ 0.2 V & self-heating law with heater off & identification of branch law \\
\addlinespace
sequence of $V_{set}$ levels: 3.0 $\to$ 0.2 V & second inter-session law and estimate of the 0.0619 V threshold & boundary identification \\
\addlinespace
sequence of $V_{set}$ levels: 0.35 $\to$ 0.06 V & paired points of both regimes and observed transition & validation of the boundary and unified balance \\
\addlinespace
seven pulses at $V_{set}=0.5$ V & resistance with heater on and independent $L$ & independent measurement of branch level \\
\addlinespace
$V_{set}=0.5$ V plateau with four heater transitions & temporal trajectory during changes in $h$ & parameter-transfer check \\
\bottomrule
\end{tabular}
\end{center}
In the last run, the parameters were not refitted and the protocol was fixed before execution. However, the form of the trajectory had been examined previously, so this is an inter-session parameter-transfer test, but not a fully independent held-out test using data that took part neither in coefficient fitting nor in model-configuration selection (Section 5.6).

\FloatBarrier
\subsection{Archive of service sessions from other magnets}
The archive was frozen on 11 August 2026 and contains 72 Siemens-magnet service sessions. In 64 of them, the CAN heater-state trace was recorded and synchronized: 49 field sessions and 15 bench sessions. Four subsets defined for distinct analysis purposes are summarized below. The shorted-node detector uses all 72 sessions; the fast-event selection in Section 6.3 uses all 56 field sessions (72 minus the bench sessions; its heater-state gates rely on the CAN trace where recorded); the remaining subsets require a CAN trace. Avanto coefficients are not transferred to other PCS devices. The field session shown in Fig. 1 is part of this same archive; the inductance of that magnet was determined from two segments of that record itself.

\begin{center}
\begin{tabular}{>{\raggedright\arraybackslash}p{1.9cm}>{\raggedright\arraybackslash}p{5.8cm}>{\raggedright\arraybackslash}p{3.9cm}l}
\toprule
Subset & Size & Inclusion criterion & Section \\
\midrule
shorted-node detector & 72 sessions $\to$ 42 episodes in 28 sessions; for comparison, 22 estimates of $R_{lead}$ from persistent-mode tails & power supply sustains voltage, current is nearly constant, no current limiting & 6.1 \\
\addlinespace
relaxation before transition to persistent mode & 21 sessions $\to$ 17 balance estimates of $R_{sw}$ on the low-voltage tail & operating current and CAN trace; decay of $V_{node}$ observed before heater switch-off & 6.2 \\
\addlinespace
archival boundary validation & 24 segments: 17 inter-session and 7 internal & constant voltage setpoint and heater off in two archived Avanto sessions & 5.4 \\
\addlinespace
fast noninductive component & 56 field sessions $\to$ 20 candidates $\to$ 18 events in 17 sessions on 12 magnets (plus one additional event outside the statistics: empty CAN structure) & fast current step at unchanged PCS state and power-supply mode; gating by mode channels and drift & 6.3 \\
\bottomrule
\end{tabular}
\end{center}
\FloatBarrier
\subsection{Processing and uncertainties}
\textbf{Time series.} The main log contains $I_{psu}$, $V_p$, $V_l$, current and voltage setpoints, and their rates; the typical sampling interval in the sessions considered is about 0.73 s. Heater state is recorded in a separate CAN trace. Before analysis, its timestamps are programmatically synchronized with the main log.

\textbf{Reconstruction and segmentation.} Node voltage is calculated using the signed subtraction $V_{node}=V_l-R_{lead}I_{psu}$. A voltage-setpoint change greater than 1 mV defines the boundary of a new constant-setpoint segment. Segments shorter than 8 s are excluded; for the remaining segments, steady-state quantities are estimated as the median over the final 40\% of the segment duration. These rules suppress transients and individual outliers.

\textbf{Relaxation before capture.} Branch resistance during the final hold is estimated by a current balance using measured $I_{psu}(t)$ as the input (Section 6.2). An estimate is accepted if the $V_{node}$ span within the window is at least 0.05 V, the branch-current contribution is at least five times the residual, and the result is stable to the addition of a linear drift term; 17 of 21 sessions with operating current and CAN trace pass these gates.

\textbf{Uncertainties.} The root-mean-square deviation of the points from the fitted law in logarithmic coordinates is 1.9--5.8\%, depending on the session; the spread of $L$ over five cycles is 0.4\%. Three estimates of $R_{path}$ agree within 0.001 m$\Omega$. For $R_{lead}$, obtained from the persistent-mode tail, the contribution of thermal drift in $V_l$ is estimated at about 0.2\%; at 500 A this corresponds to about $\pm0.5$ mA in the estimate of residual branch current. A complete uncertainty budget from measurement channels through current separation was not constructed and is listed among the limitations in Section 8.

\FloatBarrier
\section{Current Separation and Selection of Hold Duration}
\FloatBarrier
\subsection{Initial-condition problem and three measurement methods}
The coil current is not directly observed. In the reconstruction

\[
I_{coil}(t)=I_{coil}(t_0)+\frac{1}{L}\int_{t_0}^{t}V_{node}(t')\,dt',
\]
an error in the initial $I_{coil}(t_0)$ enters $I_{sw}=I_{psu}-I_{coil}$ with the opposite sign. An offset of 0.2 A is comparable with a branch current of 0.13--0.18 A and can change its estimate by roughly a factor of three. Three methods that do not depend on the unknown initial current were therefore used.

\textbf{(a) Falling edge of a pulse.} When the pulse is removed, the power-supply current decreases much faster than the current in the inductive coil can change. Immediately after the transition, the node voltage becomes close to zero, while with the heater still on the PCS branch retains a finite, nearly ohmic resistance irrespective of the vanishing Joule self-heating (Section 5.2); its current therefore also becomes nearly zero. This argument applies to $h=1$. With the heater off, $V_{node}\to0$ simultaneously removes Joule self-heating and may cause retrapping (Section 5.4). Thus, the power-supply current jump directly gives the current that flowed through the PCS before the edge:

\[
I_{sw}(t^-)\approx I_{psu}(t^-)-I_{psu}(t^+).
\]
In the actual record, the setpoint did not decrease instantaneously, but over about 4 s. During this time the coil current could still change slightly under the finite $V_{node}$. The measured jump was therefore corrected for the coil-current change

\[
\Delta I_{coil}=\frac{1}{L}\int V_{node}\,dt
\]
over the transition itself. For the 0.5 V pulses this correction was about 0.03 A; without it, PCS current would have been underestimated by about 13\%. The correction uses only measured $V_{node}$ during the transition and does not require the absolute coil current. An example is shown in Fig. 3a.

\begin{figure}[!htbp]
\centering
\includegraphics[width=\linewidth,height=0.62\textheight,keepaspectratio]{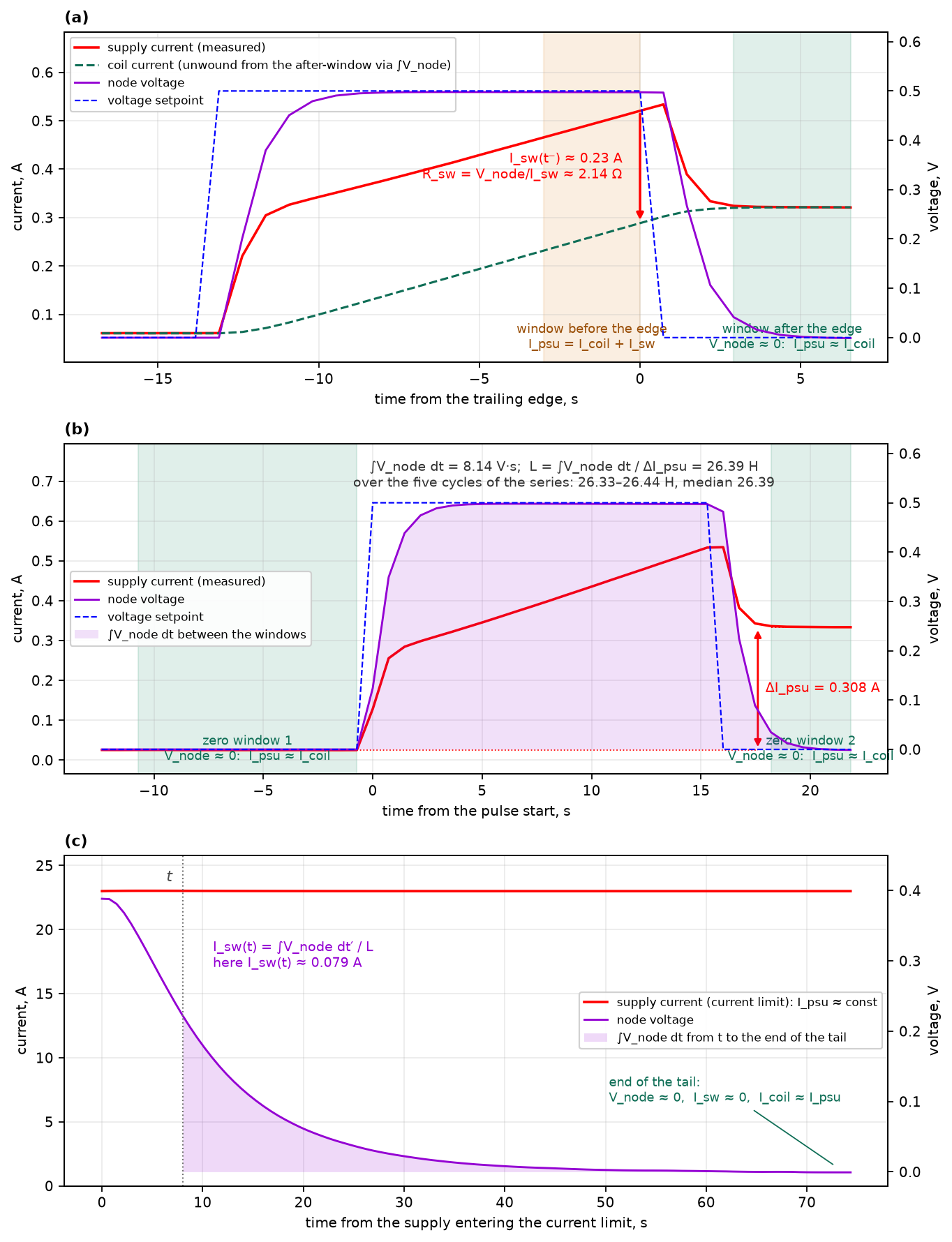}
\caption{Three methods of Section 4.1 applied to real records from the 0.5 V series described in Section 3.3. Red: power-supply current; purple: reconstructed node voltage; blue dashed: voltage setpoint. (a) Falling edge: before the edge, $I_{psu}=I_{coil}+I_{sw}$; after pulse removal, $V_{node}\approx0$ and $I_{psu}\approx I_{coil}$. The green dashed curve is the coil current taken from the post-edge window and propagated backward using the node-voltage integral. The window difference, corrected for the finite removal time, gives $I_{sw}=0.232$ A and $R_{sw}=2.14\ \Omega$, within the series ranges (0.230--0.232 A and 2.08--2.15 $\Omega$, Section 5.2). (b) Inductance from two zero-voltage windows: before and after the pulse, segments are selected where node voltage is zero within the noise and measured current equals coil current; the ratio of the node-voltage integral between the windows (shaded area) to the increase in power-supply current gives $L=26.39$ H for the cycle shown, and 26.33--26.44 H over five cycles (Section 4.2). (c) Decay under current limiting (another record from the same series, also used in Fig. 4): the power supply holds nearly constant current, and the instantaneous branch current equals the area of the remaining $V_{node}$ tail divided by $L$; the area from the selected time $t$ is shaded. At the end of the tail, $V_{node}\approx0$, $I_{sw}\approx0$, and $I_{coil}\approx I_{psu}$. None of the three methods uses the PCS characteristic.}
\label{fig:3}
\end{figure}
\textbf{(b) Inductance from two zero-voltage windows.} Two settled segments before and after the maneuver are selected in which the heater remains on, the PCS remains resistive, and the node voltage reconstructed from external measurements does not differ from zero within the noise. With the heater continuously on, the branch resistance is finite and nearly ohmic (Section 5.2), so $V_{node}\approx0$ implies $I_{sw}\approx0$, and in both windows $I_{psu}\approx I_{coil}$. For the interval between these windows, $t_1$ and $t_2$,

\[
\Delta I_{coil}=\frac{1}{L}\int_{t_1}^{t_2}V_{node}(t)\,dt,
\qquad
L=\frac{\int_{t_1}^{t_2}V_{node}(t)\,dt}{I_{psu}(t_2)-I_{psu}(t_1)}.
\]
The initial current cancels and no PCS resistance model is used. An example from the same pulse series is shown in Fig. 3b: before and after the pulse, measured current equals coil current, and the ratio of the node-voltage integral between the windows to the current increase gives the inductance.

\textbf{(c) Decay under current limiting.} When the power supply is already holding a nearly constant total current, subsequent redistribution occurs between the coil and PCS. The physical basis of the method is that, once the total current is fixed, all current still flowing through the PCS at the present moment must eventually move into the coil by the end of the relaxation. The increase of coil current from the present time to the end of the transient equals the integral of the remaining node voltage divided by $L$. Therefore, this area is the current presently flowing through the PCS:

\[
I_{sw}(t)=\frac{1}{L}\int_t^{\infty}V_{node}(t')\,dt'.
\]
In other words, $I_{sw}$ is determined by the area under the remaining $V_{node}$ tail; a real-record example is shown in Fig. 3c. In practice, integration ends when the residual decay is no longer distinguishable from noise. By that time, current through the PCS is negligible and the coil current has nearly stopped changing; consequently $V_{node}\approx0$, and the current-lead resistance can be estimated independently as $R_{lead}=V_l/I_{psu}$. The low current used in these maneuvers simplifies the experiment and reduces its energy scale, but the method itself is not mathematically dependent on small current.

\FloatBarrier
\subsection{Direct measurement of inductance}
Of the three methods in Section 4.1, only method (b) is used here for direct determination of $L$. Methods (a) and (c) are used mainly to determine branch current; method (c) also gives $R_{lead}$ at the end of the tail.

Across five cycles, method (b) gives $L=26.33/26.39/26.40/26.44/26.33$ H: median 26.39 H, full range 0.4\%. The measurement uses neither the PCS characteristic nor a model of the external path.

\FloatBarrier
\subsection{Coil-current matching and settling time}
In persistent mode, the coil current is not observable at the external terminals. After the heater is turned on, the PCS becomes resistive and

\[
I_{psu}=I_{coil}+I_{sw}.
\]
On a settled segment with $V_{node}\approx0$, branch current is close to zero, so

\[
I_{coil}\approx I_{psu}.
\]
Such a window provides an independent reference point for $I_{coil}$. Equality is not established instantaneously: immediately after switching, part of the current still flows through the PCS, so a hold is required. The causal relation is as follows: as long as $I_{coil}\neq I_{psu}$, their difference must flow through the resistive branch, and therefore a nonzero voltage $V_{node}=R_{sw}I_{sw}$ must exist at the node; it is this voltage that transfers current from the branch into the coil. Hence $V_{node}\to0$ is the result of completed relaxation, not its initial condition.

For a resistive PCS,

\[
V_{node}=L\frac{dI_{coil}}{dt}=R_{sw}I_{sw},
\qquad
I_{sw}=I_{psu}-I_{coil}.
\]
Thus the general mismatch equation is

\[
\frac{dI_{sw}}{dt}+\frac{R_{sw}}{L}I_{sw}=\frac{dI_{psu}}{dt}.
\]
When $I_{psu}$ is practically constant and the heater is on, so that the PCS remains in the low-voltage resistive regime and $R_{sw}$ changes little, branch current decays exponentially:

\[
I_{sw}(t)=I_{sw}(0)\exp\left(-\frac{t}{\tau}\right),
\qquad
\tau=\frac{L}{R_{sw}}.
\]
If the tolerance is specified as a relative coil-current error $\varepsilon$, the required hold is

\[
t_{hold}\ge \tau\ln\frac{|I_{sw}(0)|}{\varepsilon|I_{coil}|}.
\]
A hold of $3\tau$ leaves 5\% of the initial mismatch, while $4.6\tau$ leaves 1\%. For the Avanto, $L=26.39$ H and $R_{sw}\approx2.0\ \Omega$ give $\tau\approx13$ s and $3\tau\approx40$ s. There is therefore no universal hold time: it depends on $L$, the PCS state, the initial mismatch, and the allowed error.

The exponential approximation applies while $I_{psu}$ is practically constant, the heater is on, the PCS remains resistive, and $R_{sw}$ changes little; under these conditions $V_{node}$ decays to zero together with $I_{sw}$. Otherwise, the full equation including $dI_{psu}/dt$ is required.

\FloatBarrier
\subsection{Validation on Avanto data and reference-point options}
In one session, an early 4.4 s window showed a deficit of 0.263 A, which decreased to 0.041 A after 26 s. With $\tau=13.2$ s, the model predicts 0.037 A; this single nontrivial point agrees with the exponential to within about 12\%, but it is not a precision validation of the exponential. The value 0.263 A itself sets the scale of the model and therefore is not an independent check. Both reference windows and the model exponential are shown in Fig. 4.

\begin{figure}[!htbp]
\centering
\includegraphics[width=\linewidth]{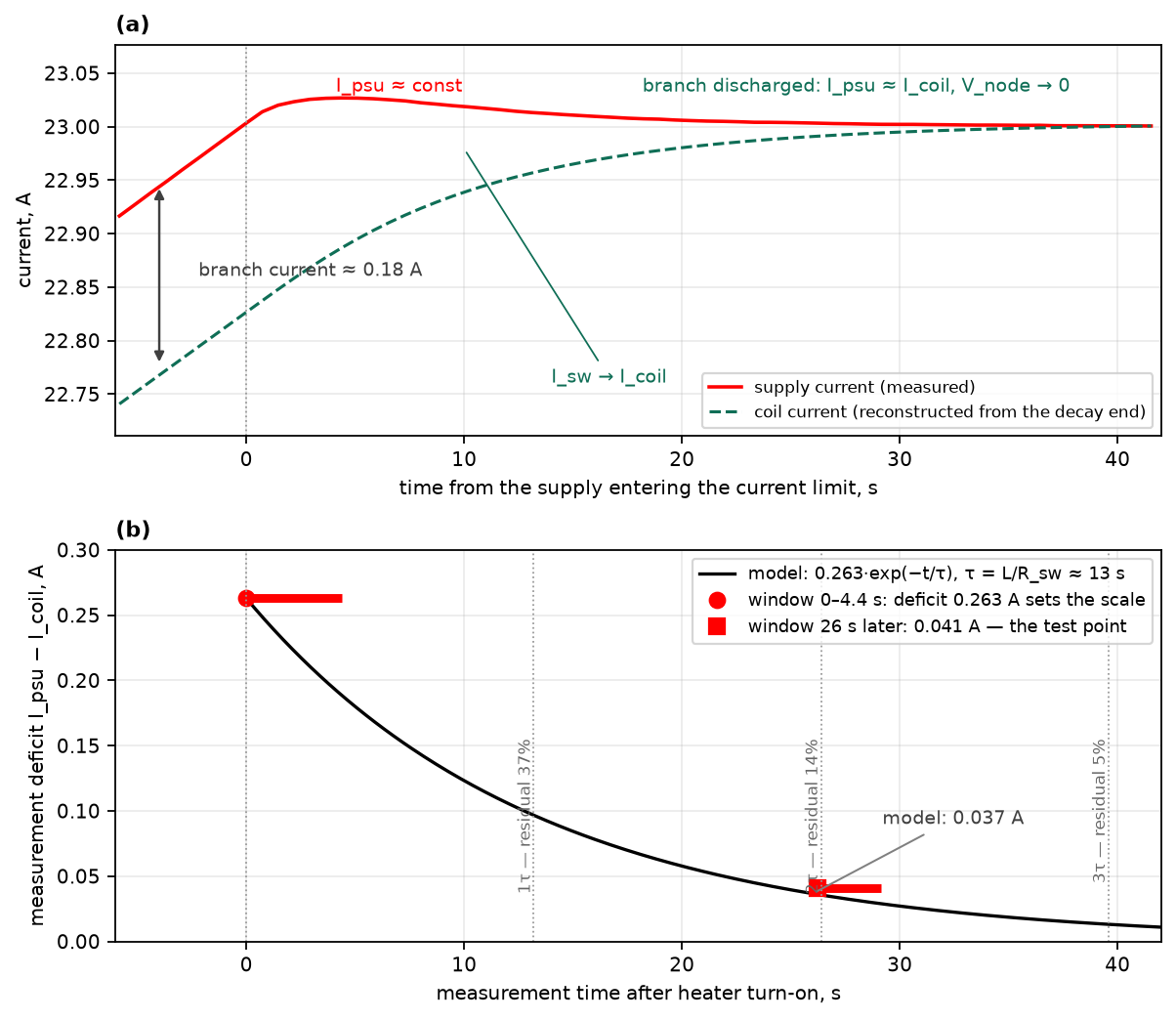}
\caption{Coil-current matching. (a) At practically constant $I_{psu}$, PCS branch current (about 0.18 A) transfers into the coil and the reconstruction of $I_{coil}$ converges to the measured current. (b) Over 26 s, the deficit decreases from 0.263 to 0.041 A; the model $0.263\exp(-t/\tau)$ with $\tau\approx13$ s gives 0.037 A. Dashed markers indicate the remaining 37\%, 14\%, and 5\% after $\tau$, $2\tau$, and $3\tau$, respectively.}
\label{fig:4}
\end{figure}
The 0.263 A deficit is comparable with the branch current of 0.13--0.18 A, so an excessively early reference window changed its estimated value by roughly a factor of three.

If the initial coil current is unknown, a small setpoint and a hold until $V_{node}\approx0$ were used after opening the PCS. At this point, the PCS current has practically vanished and $I_{coil}\approx I_{psu}$, providing a direct reference point for coil current. In three episodes, this estimate differed from the reconstruction of $I_{coil}$ obtained by integrating $V_{node}/L$ by 0.08/0.25/0.15 A at currents of 23/31/34 A, i.e. only 0.3--0.8\%. This regulator behavior is specific to the power-supply configuration used here: transfer of the method to another installation requires separate verification that the power supply permits this operating mode and correctly measures the resulting current.

If the coil current is known approximately, the power-supply current can be preset close to it. This reduces the initial PCS current and shortens the required hold before $I_{psu}$ is used as an estimate of $I_{coil}$. Such a reference point also allows accumulation of error in the integral reconstruction to be monitored; in one 900 s maneuver, the discrepancy reached 1.3\%.

\FloatBarrier
\section{Quantitative Characterization of the Siemens MAGNETOM Avanto PCS}
All results in Section 5 were obtained on a single PCS of a Siemens MAGNETOM Avanto magnet in dedicated maneuvers.

\FloatBarrier
\subsection{Initiation and persistence of the resistive state}
In the superconducting state, the PCS shunts the node, so $V_{node}\approx0$ and electrical dissipation in its branch is negligible. In the maneuvers studied, transition of the PCS into the resistive state was observed only after the heater was turned on; electrical initiation of the transition by applied voltage was not observed. Once the resistive state had formed, finite $V_{node}$ produced internal dissipation $V_{node}^2/R_{sw}$, and the state could persist after heater switch-off.

The distinction between initiation and persistence is seen directly in the experiment. During a sequence of voltage levels, the voltage setpoint $V_{set}$ was reduced to zero for four seconds and then returned to the same 0.10 V. Before the zero interval, this setpoint sustained the resistive state; afterward, $V_{node}$ remained near zero and the power-supply current reached the imposed limit. It is important that what is identical in the two episodes is the external setpoint $V_{set}$, not the voltage across the PCS. In the resistive state, part of the external voltage is indeed applied to the branch, whereas after retrapping the superconducting branch shorts the node and keeps $V_{node}\approx0$ under the same external command of 0.10 V. Thus, the same external setpoint is compatible with two PCS states, selected by the system history; both episodes are shown in Fig. 5.

\begin{figure}[!htbp]
\centering
\includegraphics[width=\linewidth]{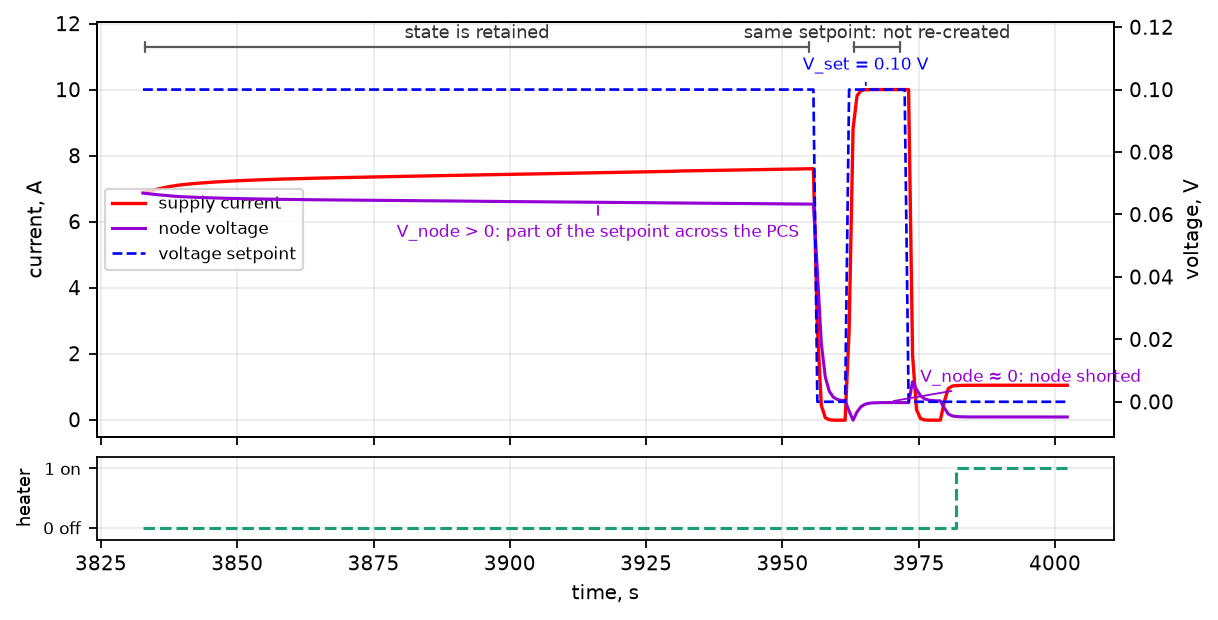}
\caption{A single 0.10 V setpoint sustains an existing resistive state but does not restore it after a four-second zero-voltage interval. The lower trace is heater state (1 = on, 0 = off): both compared episodes occur at $h=0$; the heater is turned on again only at the end of the window.}
\label{fig:5}
\end{figure}
At a 0.10 V setpoint, both PCS states are possible in the observed maneuver. The state is determined by history, while transition into the resistive state is initiated by the heater.

This establishes the procedural role of the heater: it initiates the resistive state, but the subsequent existence of that state does not coincide in time with $h=1$. At sufficient node voltage, the state can be sustained by its own dissipation (Sections 5.3--5.4).

Additional evidence was obtained in another dedicated session of the same Avanto: for 59 s after heater switch-off, the electrical dynamics remained continuous and no sign of PCS return to the superconducting state was observed. Therefore, the recorded heater state cannot be used as a direct time label for the PCS state.

\FloatBarrier
\subsection{Heater on: low-voltage ohmic regime}
Three measurement methods that do not require an assumed initial coil current were applied in two sessions:

\begin{center}
\begin{tabular}{>{\raggedright\arraybackslash}p{5.4cm}l>{\raggedright\arraybackslash}p{4.0cm}}
\toprule
Method & Node-voltage window & Result \\
\midrule
seven pulses at a 0.5 V setpoint, six accepted into the statistics & about 0.5 V & 2.08--2.15 $\Omega$ (branch current 0.230--0.232 A) \\
\addlinespace
nine constant $V_{set}$ levels & 0.05--0.35 V & 1.86--1.99 $\Omega$, with no pronounced slope \\
\addlinespace
decay in current-limit mode & 0.39 $\to$ 0.013 V & 2.19 $\Omega$, ohmic continuation toward low voltage \\
\bottomrule
\end{tabular}
\end{center}
Within each low-voltage series, the resistance is consistent with an ohmic regime; the combined inter-session range of estimates is 1.86--2.19 $\Omega$. This does not imply that the resistance remains constant at arbitrary voltage.

The low-voltage plateau is consistent with heater input dominating PCS self-heating: the joint fit in Section 5.5 gives an effective heater power $P_h=2.20$ W, whereas at 0.35 V the electrical power of the branch is only about 0.063 W. This is an interpretation within the model, not a direct calorimetric measurement.

In the pulse series on one PCS, no systematic dependence on heater-on duration was detected: accepted measurements span 109--819 s; branch current changes from 0.230 to 0.232 A, less than 1\%, while the calculated resistance changes from 2.08 to 2.15 $\Omega$, or 3.3\%. The difference between these relative ranges results from small variations of the actual $V_{node}$. The observed rise of voltage to its plateau took about 3 s, but the first pulse suitable for a resistance estimate was obtained only after 109 s. These data therefore do not determine a shorter thermal settling time of the PCS (Fig. 6).

\begin{figure}[!htbp]
\centering
\includegraphics[width=\linewidth]{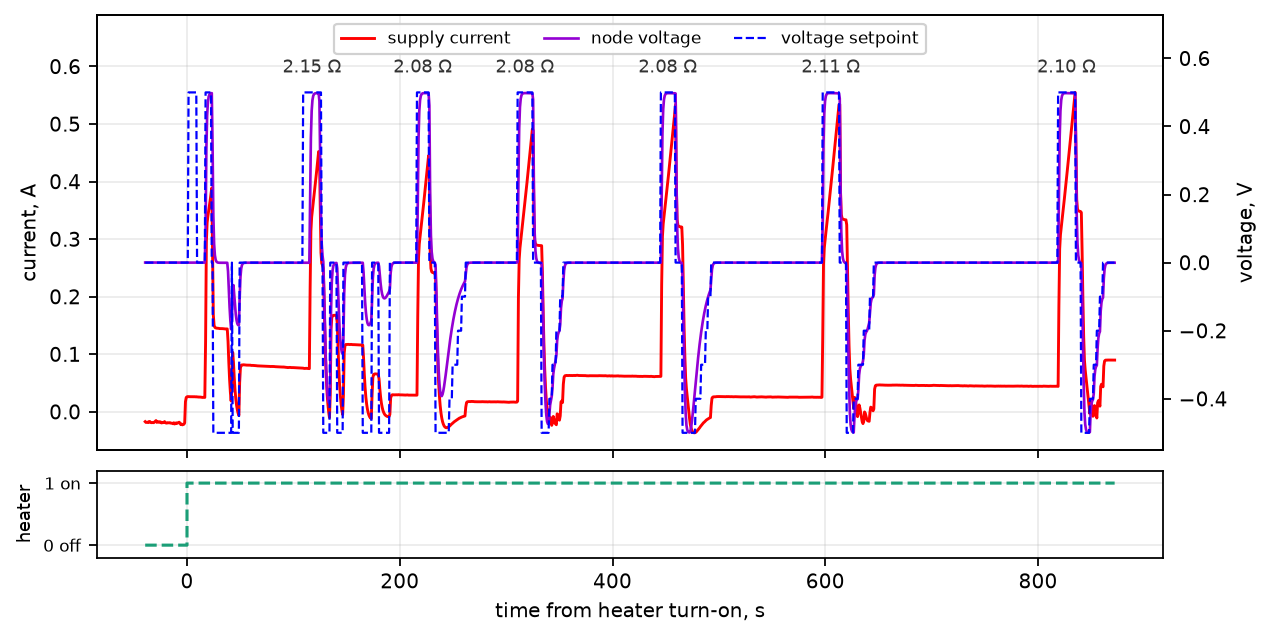}
\caption{0.5 V pulses with the heater continuously on. Over 109--819 s, the resistance is 2.08--2.15 $\Omega$ with no visible time trend. The first pulse was preceded by a long zero-voltage interval needed for an independent determination of the coil current and is not included in the statistics.}
\label{fig:6}
\end{figure}
\FloatBarrier
\subsection{Heater off: self-heating and the power law}
With the heater off, reconstructed steady values of $R_{sw}$ increased systematically with $|V_{node}|$. In logarithmic coordinates, the points formed an almost straight line, so a power law $R_{sw}\propto |V_{node}|^n$ was tested as the minimal two-parameter description. The same form was independently reproduced in three sessions using different voltage-change protocols. For comparison between sessions, the law is written in dimensionless normalization as

\[
\frac{R_{sw}}{1\ \Omega}
=k\left(\frac{|V_{node}|}{1\ \text{V}}\right)^n.
\]
The numerical value of $k$ therefore corresponds to the resistance in ohms when the numerical value of $V_{node}$ is expressed in volts.

\begin{center}
\begin{tabular}{lrrrr}
\toprule
$V_{set}$ protocol & Range & $k$ & $n$ & RMS deviation in log coordinates \\
\midrule
smooth decrease 1.0 $\to$ 0.2 V & 0.067--0.822 V & 0.655 & 1.51 & 1.9\% \\
sequence of levels 0.35 $\to$ 0.06 V & 0.025--0.322 V & 0.648 & 1.50 & 5.8\% \\
sequence of levels 3.0 $\to$ 0.2 V & 0.33--2.4 V & 0.640 & 1.44 & 3\% \\
\bottomrule
\end{tabular}
\end{center}
Both regimes measured at the same voltage levels within one session are shown in Fig. 7.

\begin{figure}[!htbp]
\centering
\includegraphics[width=\linewidth]{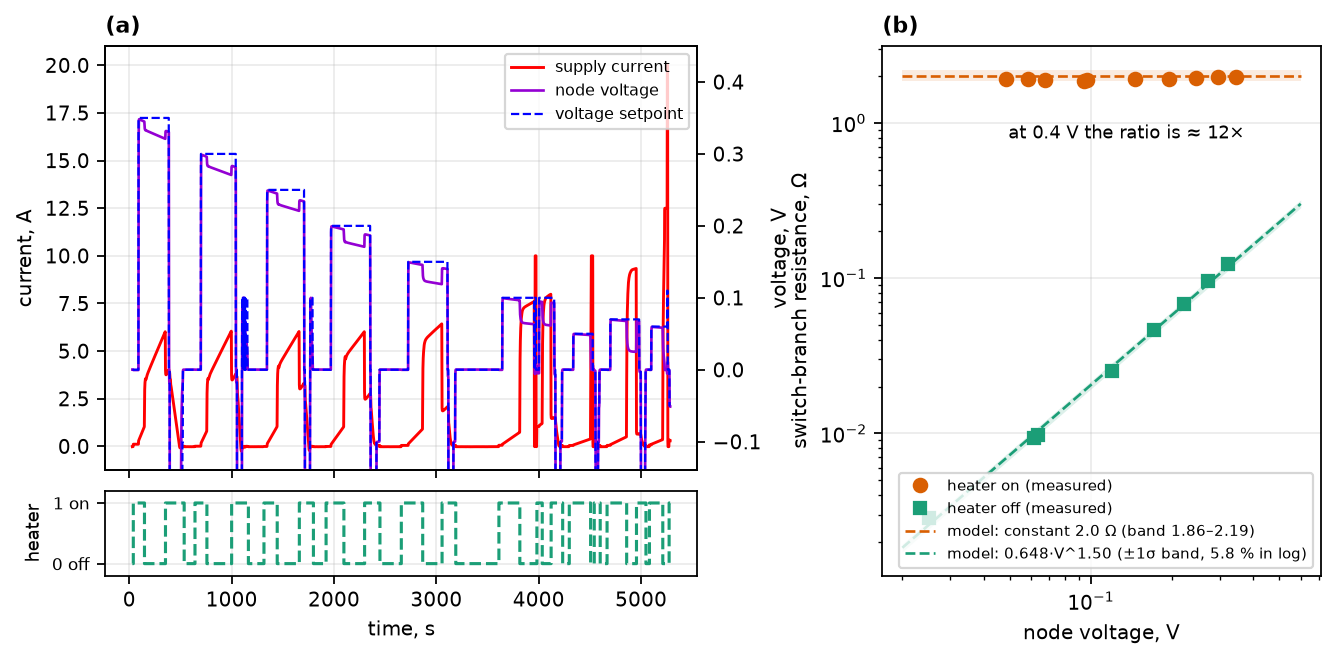}
\caption{Two resistive regimes at the same voltage levels. (a) Maneuver from which the points were extracted: the setpoint is reduced in steps from 0.35 to 0.06 V and the heater is switched at each level (lower trace: heater state); auxiliary recharge intervals between levels extend below the lower edge of the voltage scale. (b) Steady-state points: nearly constant resistance at $h=1$ and a power-law dependence at $h=0$.}
\label{fig:7}
\end{figure}
Comparison of the two regimes at the same voltage gives the scale of the effect: at 0.4 V, the resistances are about 2.0 $\Omega$ and 0.164 $\Omega$, respectively, differing by about a factor of twelve.

Across the three sessions with different protocols, $k$ lies between 0.640 and 0.655 and $n$ between 1.44 and 1.51. The full range of $k$ is 2.3\% relative to its minimum value. The closeness of the estimates supports reproducibility of the power-law description for this PCS over the overlapping measured ranges, but does not establish universality of a single law outside those windows. The data used for this law --- the setpoint decrease and conversion of steady segments into branch resistance --- are shown in Fig. 8.

\begin{figure}[!htbp]
\centering
\includegraphics[width=\linewidth]{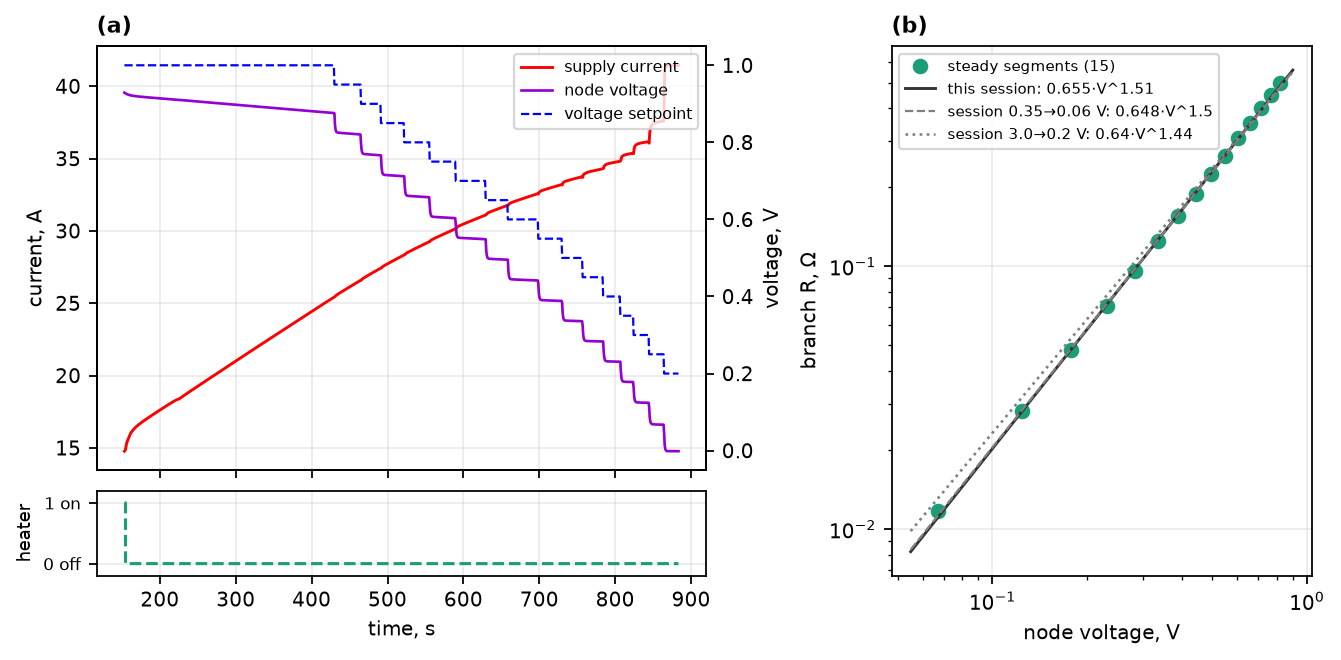}
\caption{Power-law regime with the heater off. (a) Setpoint decrease from 1.0 to 0.2 V: each change is followed by a segment with nearly constant setpoint, while the power-supply current continues to increase because the coil integrates the node voltage. (b) Fifteen steady segments converted to branch resistance; $I_{sw}$ is obtained by subtracting reconstructed $I_{coil}$ from measured $I_{psu}$, using an initial condition set when the coil was nearly empty. Gray lines show the laws from the two other sessions.}
\label{fig:8}
\end{figure}
\FloatBarrier
\subsection{Self-sustain boundary as a joint property of the PCS and the external circuit}
With the numerical form

\[
R_{sw}[\Omega]=k\bigl(|V_{node}|[V]\bigr)^n,
\]
the branch current is

\[
I_{sw}=\frac{|V_{node}|^{1-n}}{k}.
\]
Because $n>1$, reducing $V_{node}$ increases the branch current. This would appear paradoxical for a fixed resistor, but the PCS is not a fixed resistor: when voltage decreases, Joule heating decreases, the normal zone cools, and its resistance falls faster than the voltage itself; as a result, the ratio $I_{sw}=V_{node}/R_{sw}(V_{node})$ increases. The measured characteristic shows this numerically: at $V_{node}=0.4$ V, branch current is about 2.4 A, whereas at 0.025 V it is already 8.76 A (observation below). A sixteen-fold lower voltage therefore drives about three and a half times more current through the branch. This creates positive feedback: lower $V_{node}$ $\to$ less heating $\to$ lower $R_{sw}$ $\to$ larger $I_{sw}$ $\to$ larger drop across $R_{path}$ $\to$ still lower $V_{node}$.

It is useful to state explicitly what must be self-consistent before writing the boundary formula. (1) For an assumed $V_{node}$, the PCS characteristic determines the required $I_{sw}$. (2) This current contributes to the total current in the external path and creates an additional drop across $R_{path}$. (3) The external circuit therefore determines, in turn, what $V_{node}$ actually remains at the node. (4) An operating point exists only where both conditions are simultaneously satisfied --- where the branch characteristic intersects the external-circuit load line. (5) As the available voltage decreases, the two solutions approach one another and merge at a tangency point. (6) Below this tangency, there is no self-consistent resistive solution and retrapping is observed.

Strictly speaking, the operating point is quasi-steady. When $V_{node}\neq0$, coil current continues to change slowly according to $L\,dI_{coil}/dt=V_{node}$, so there is no absolute static equilibrium. However, the PCS thermal state settles on a much shorter time scale, and on that scale $I_{coil}$ is treated as an approximately fixed parameter. The operating point discussed below is therefore a quasi-steady PCS operating point. Graphically, within the identified power-law model, it is defined by the intersection of the PCS characteristic and the load line (Fig. 9a).

\begin{figure}[!htbp]
\centering
\includegraphics[width=\linewidth]{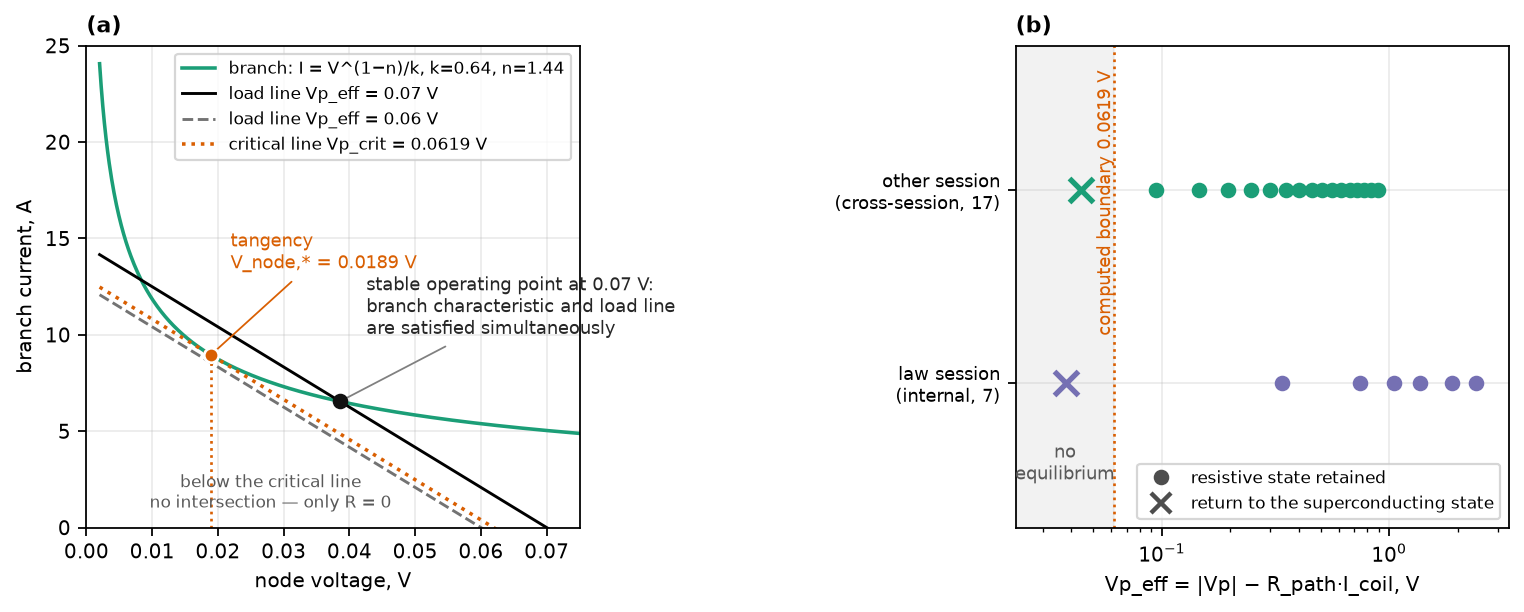}
\caption{Self-sustain boundary in the identified model. (a) Branch characteristic from another session ($k=0.640$, $n=1.44$) and load lines of the external path: at $V_{p,eff}=0.070$ V an intersection exists; at 0.060 V it does not. The critical line $V_{p,crit}=0.0619$ V is tangent to the characteristic at $V_{node,*}=0.0189$ V. (b) Locations of 17 inter-session and seven internal segments relative to this boundary; circles indicate persistence of the resistive state and crosses indicate return to the superconducting state.}
\label{fig:9}
\end{figure}
The boundary condition is

\[
V_{p,eff}\ge V_{p,crit}
=
\left[
\frac{R_{path}\,n^n}
{k(n-1)^{n-1}}
\right]^{1/n},
\qquad
V_{p,eff}=|V_p|-R_{path}|I_{coil}|.
\]
The boundary is derived for the observed polarity, in which $V_p$, $I_{coil}$, and $V_{node}$ have the same sign. Once that branch is fixed, the magnitude notation is simply a compact form of the signed relationship in Section 3.2. For the opposite polarity, the original signed equation must be used; replacing the variables by magnitudes is not permissible. At the tangency point, the node voltage is not equal to the critical power-supply voltage but is

\[
V_{node,*}=\frac{n-1}{n}V_{p,crit}.
\]
Once the operating point is lost, self-heating within the model can no longer sustain the normal zone; in the observed maneuvers, the PCS then returned to the superconducting state. Because the observed boundary depends not only on branch parameters $k$ and $n$, but also on the external-path resistance and coil current, it is determined by the entire connected system and is not solely an internal PCS parameter.

\textbf{Observation.} At a 0.07 V setpoint, the resistive state persists ($V_{node}=0.025$ V, $I_{sw}=8.76$ A, branch power 0.219 W); at the next setpoint of 0.06 V, the PCS returns to the superconducting state.

\textbf{Calculation.} $R_{path}=4.802$ m$\Omega$ is determined directly from episodes with $V_{node}\approx0$. The law from another session ($k=0.640$, $n=1.44$) gives $V_{p,crit}=0.0619$ V --- an inter-session estimate lying between the observed levels 0.06 and 0.07 V. The corresponding node voltage at tangency is $V_{node,*}=0.0189$ V. The law from the same session as the observed transition ($k=0.648$, $n=1.50$) gives 0.0718 V, 2.6\% above the 0.07 V setpoint that still sustained the resistive state; this discrepancy characterizes the accuracy with which the power-law approximation localizes a transition resolved only at discrete setpoint levels. The inter-session estimate of 0.0619 V, which lies between the two observed outcomes, provides the stronger cross-session evidence.

\textbf{Retrospective validation.} In two archived sessions, the criterion separates the outcomes of all 24 constant-voltage-setpoint segments: the resistive state persisted in 22 and the PCS returned to the superconducting state in two. The effective setpoint includes the drop $R_{path}I_{coil}$; at 32--34 A, this correction is about 0.16 V, and without it the outcomes would have been misclassified. For 17 segments of the 1.0 $\to$ 0.2 V law, branch parameters and $R_{path}$ were determined from other sessions; this is the inter-session part of the validation. The remaining seven segments belong to the session used to determine the threshold law itself and therefore check only internal consistency. Figure 9 shows the boundary and the positions of all 24 segments relative to it.

The observed signature of return to the superconducting state is that $V_{node}$ falls approximately to zero under a nonzero setpoint while power-supply current reaches the level imposed by the external path. The temporal form of such an approach to the boundary is shown by the illustrative minimal node-model trajectory in Section 6.2 (panel (b) of the illustrative figure).

\FloatBarrier
\subsection{Effective electrothermal description of the two regimes}
Both measured resistive regimes can be described by a single phenomenological electrothermal relation. Its status should be stated explicitly: it is not a dynamic thermal model of the PCS and not a reconstruction of PCS temperature; it is a minimal algebraic description of steady operating points. The power-law form was chosen empirically from the data; the thermal balance below shows why such a dependence arises naturally. If effective heat removal is linear in the temperature rise $\Delta T$,

\[
G\Delta T=P_hh+\frac{V^2}{R},
\]
and the resistance of the resistive region grows as

\[
R=R_0(\Delta T)^a,
\]
then eliminating the unobserved $\Delta T$ gives

\[
R=C\left(P_hh+\frac{V^2}{R}\right)^a,
\qquad
C=\frac{R_0}{G^a},
\]
where $h\in\{0,1\}$ is the recorded heater state. It is used as an observed control input and is not a measurement of heater current, temperature, or actual heater power. Since zone temperature is not measured, $R_0$ and $G$ cannot be identified separately from these data; only their combination $C$ is used below. The fit is performed in the same dimensionless normalization as in Section 5.3,

\[
\frac{R}{1\ \Omega}
=C\left(
\frac{P_hh+V^2/R}{1\ \text{W}}
\right)^a,
\]
so the numerical value of $C$ is dimensionless. The relation is static: the balance is written for steady states, without heat capacity or settling time, and all points included in the fit are steady segments. The temporal evolution of transients --- settling after a pulse (Section 4.1) or changes in effective resistance along the final hold (Section 6.2) --- is not described by this relation; individual states along a transient can be compared with it only in a quasi-steady sense.

Fitting paired points from one session, in which both regimes were measured at the same voltage levels (10 points at $h=1$, 8 at $h=0$), gives

\[
C=0.1773,\qquad a=3.00,\qquad P_h=2.20\ \text{W}.
\]
The RMS deviation in logarithmic coordinates is 4.1\%. A single three-parameter relation describes resistance from 0.0029 to 1.99 $\Omega$, a 690-fold range (Fig. 10).

\begin{figure}[!htbp]
\centering
\includegraphics[width=\linewidth]{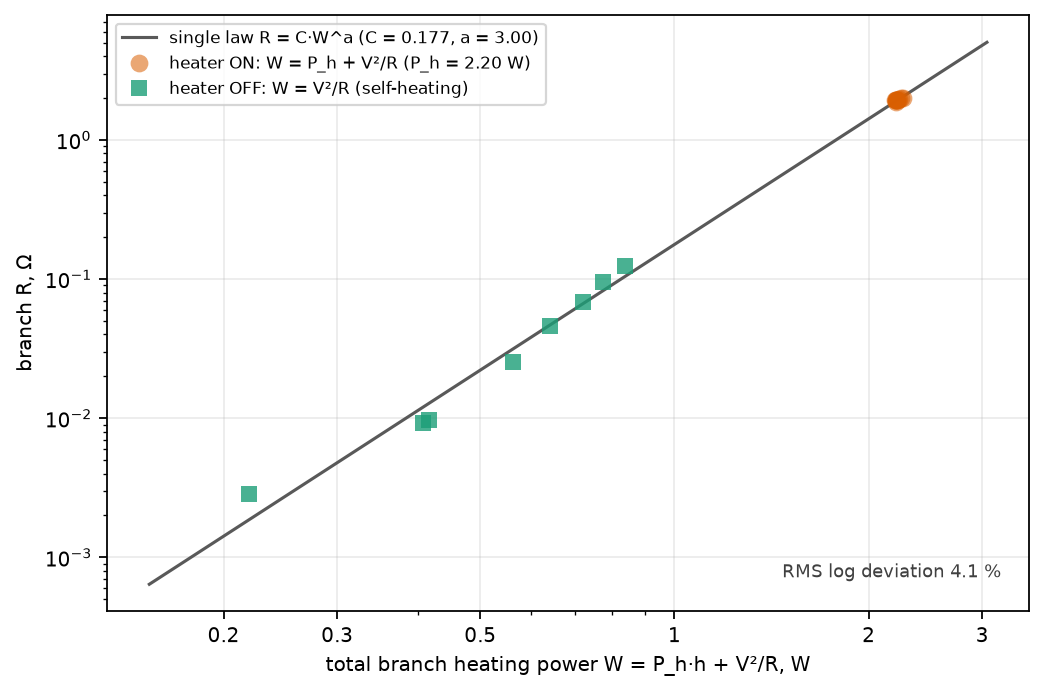}
\caption{The two regimes in coordinates of resistance and effective power $W=P_hh+V^2/R$. All 18 points are consistent with a single law $R=CW^a$ over a 690-fold range of resistance. The ten points at $h=1$ lie within one marker size and visually overlap.}
\label{fig:10}
\end{figure}
The fitted parameters are consistent with both original branches of the model. For $h=0$, the balance reduces to $R=C(V^2/R)^a$, so $R^{1+a}\propto V^{2a}$ and $R\propto V^{2a/(1+a)}$. Thus the power-law exponent is

\[
n=\frac{2a}{1+a};
\]
$a=3.00$ gives $n=1.50$, whereas direct fits to individual sessions give $n=1.44-1.51$. For $h=1$, the model gives the limit

\[
R(V\to0)=CP_h^a=1.89\ \Omega,
\]
close to the measured low-voltage plateau of 1.86--1.99 $\Omega$, and predicts only a small increase of resistance with voltage. Because these comparisons use the same data from which $C$, $a$, and $P_h$ were fitted, they test internal model consistency rather than providing an independent validation.

The exponent $a$ is determined to a large extent by the power law of the $h=0$ regime, while $P_h$ is determined by the resistance level at $h=1$. These parameters should therefore not be regarded as independently established internal physical characteristics. The model does not identify temperature, heat-transfer coefficient, size, or location of the resistive region. $P_h$ is only an effective heater contribution within the chosen balance, not a nameplate or directly measured heater power.

\textbf{PCS state as a history variable.} The characteristic $R(V,h)$ applies only when the branch is already resistive; by itself, it does not define the PCS state. The state logic is shown in Fig. 11 and reduces to three rules: turning on the heater initiates the resistive state; at $h=0$, the resistive state persists as long as the external-circuit load line intersects the self-heating characteristic (criterion of Section 5.4); when the operating point is lost, retrapping occurs and the branch returns to the superconducting state, shorting the node. If the heater is switched off when the system is already below the boundary, the model has no resistive operating point and the PCS transitions directly to the superconducting state. In the maneuvers examined, self-heating was observed as a mechanism that maintained a previously created normal zone; spontaneous transition from the superconducting state at $h=0$ was not observed.

\begin{figure}[!htbp]
\centering
\includegraphics[width=\linewidth]{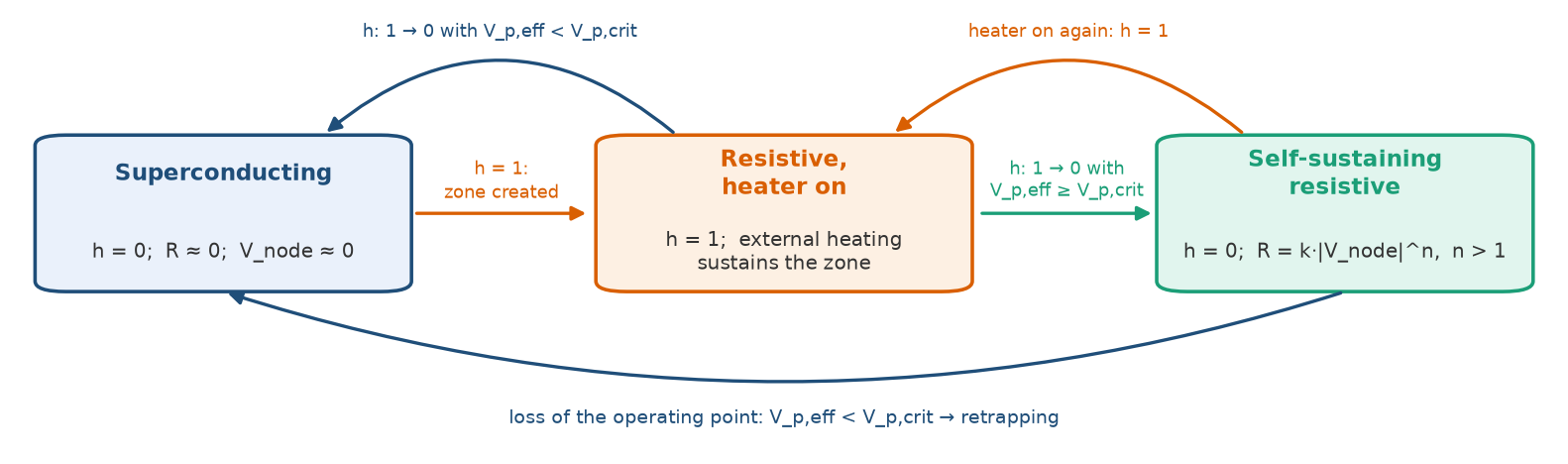}
\caption{PCS state logic. The $R(V,h)$ characteristic defines resistance only in the two resistive states; loss of the self-heated operating point causes retrapping.}
\label{fig:11}
\end{figure}
In the maneuvers considered, the state trajectory is known, so the recorded $h$ together with the threshold rule of Section 5.4 is sufficient. For processing the archival corpus, the same rule is used in a simplified threshold form: an $h=0$ interval bounded by two heater-on events is treated as resistive if the median $|V_{node}|$ on that interval is not below the threshold; intervals before the first heater-on event and after the last are always treated as superconducting. A model of an arbitrary ramp must track the PCS state explicitly; a complete state machine is not validated in the present work (Section 8).

\FloatBarrier
\subsection{Inter-session reproduction of dynamics with fixed parameters}
The dynamic check was performed on a 0.5 V plateau from another session of the same PCS containing four heater transitions (Fig. 12). The purpose was to determine whether it is sufficient merely to include a finite PCS branch resistance or whether the two regimes must be distinguished explicitly through heater state $h$. The pair of characteristics $R(V,h)$ --- the nearly ohmic regime at $h=1$ and the power law at $h=0$ --- had been determined from other sessions, and its coefficients were not changed for this run.

\begin{figure}[!htbp]
\centering
\includegraphics[width=\linewidth]{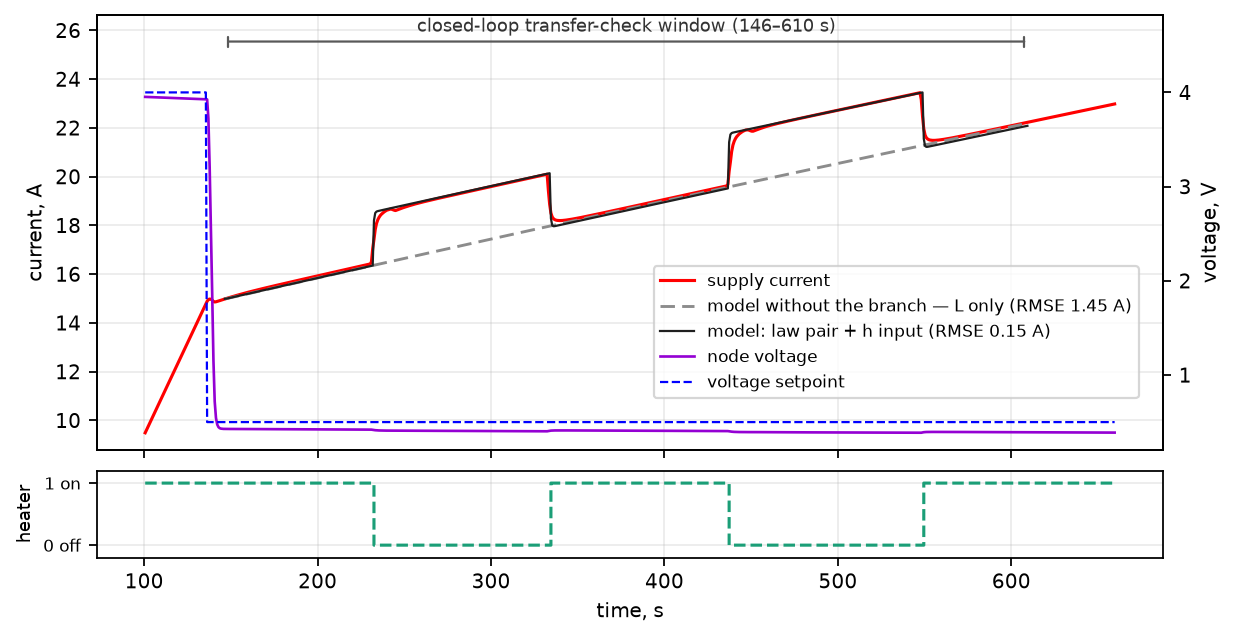}
\caption{Inter-session reproduction of a 0.5 V plateau with four heater transitions. The fixed $R(V,h)$ pair reproduces the current jumps and slopes (RMSE 0.154 A), whereas the model without a PCS branch does not (1.449 A). The bracket marks the 146--610 s window over which RMSE is calculated.}
\label{fig:12}
\end{figure}
Three alternatives were used for comparison. The first contains no PCS branch at all. In the second, the PCS is represented by a constant resistance $R_{sw}=2.0\ \Omega$, corresponding to the measured low-voltage level with the heater on. The third uses a single law

\[
R_{sw}=2.258|V_{node}|^{0.99},
\]
independent of heater state. This law was obtained from one common fit to all 18 paired points from the calibration staircase, the session in which both regimes were measured at the same voltage levels (Fig. 7). Thus, both alternatives were parameterized before the test run: the single law is the optimal fit to all 18 points, while the constant 2.0 $\Omega$ value is a representative measured level of the ohmic regime; nothing was fitted to the test session.

The calculation window begins after the rising edge into the plateau and covers 146--610 s. The initial $V_{node}=0.43$ V is taken from the first sample in the window, without using future data. Each model is initialized with a branch current consistent with that model; for the $R(V,h)$ pair, the initial branch current is 0.215 A. This protocol excludes adjustment of the initial state using the subsequent trajectory.

\begin{center}
\begin{tabular}{>{\raggedright\arraybackslash}p{7.2cm}r>{\raggedright\arraybackslash}p{3.7cm}}
\toprule
Branch configuration & Current RMSE & Relative to model without branch \\
\midrule
no switch branch & 1.449 A & 1.00$\times$ \\
\addlinespace
fixed two-regime model: ohmic regime ($h=1$) + power law ($h=0$) & 0.154 A & 9.4$\times$ \\
\addlinespace
constant $R_{sw}$ without $h$ input & 1.456 A & 1.00$\times$ \\
\addlinespace
single $R(V)$ law without $h$ input & 1.449 A & 1.00$\times$ \\
\bottomrule
\end{tabular}
\end{center}
In this maneuver, the model pair with explicit $h$ input reduces RMSE by a factor of 9.4, while both fixed alternatives without $h$ are practically indistinguishable from the model without a PCS branch. The reason is straightforward: a branch carrying nearly constant current merely offsets $I_{psu}$ from $I_{coil}$ by a constant and cannot create the observed current jumps when the heater is switched; the information lies precisely in those jumps and slopes. The agreement of the single-law model with the no-branch model to three decimal places is not accidental: the exponent 0.99 means $R_{sw}\propto|V_{node}|$, i.e. a branch with nearly constant current $I_{sw}=V_{node}/R_{sw}\approx0.44$ A (0.43--0.44 A over 0.1--0.5 V). After causal initialization, such a branch only shifts the coil current by a constant amount and leaves the power-supply current trajectory the same as in the no-branch model. A single curve drawn through both regimes therefore degenerates into a constant-current branch with no mechanism for redistribution when the heater switches.

Thus, the improvement is associated not with the mere presence of a resistive branch, but with separating the two regimes. For dynamic modeling of this PCS, the recorded heater state must enter the model explicitly; neither a single constant nor a single $R(V)$ curve separates the two regimes.

The plateau from this session had already been analyzed in the first revision of the work and contributed to selection of the two-regime configuration itself. The final run is therefore an inter-session reproduction with fixed coefficients and causal initialization, but not a fully independent held-out test. Independent validation would require a second maneuver that had not been examined in advance; no such data set is available in the present work.

\FloatBarrier
\section{Supporting Evidence from Field Logs of Other Siemens Magnets}
Section 5 established numerical laws for the PCS of the Avanto studied here. The narrower question in this section is whether previously recorded logs from other Siemens magnets contain observable consequences of the same electrical topology and current-redistribution dynamics. This section does not test transferability of the Avanto coefficients and does not validate its numerical boundary on other PCS devices; doing so would require dedicated maneuvers on each switch.

\FloatBarrier
\subsection{Transferable consequences of the circuit at operating current}
Signed reconstruction of $V_{node}$ and the associated asymmetry between ramp-up and ramp-down were introduced in Section 3.2. Only the consequence is recalled here: the voltage drop across the external path reduces $|V_{node}|$ during ramp-up and increases it during ramp-down. Therefore, within the criterion of Section 5.4, approach to the self-sustain boundary is most likely during the final approach to the target current, when the power-supply voltage is decreasing while the current is still large. The numerical Avanto boundary is not transferred to other magnets here.

A search using observable electrical signatures --- the power supply maintains nonzero voltage, current is practically constant, and no current limiting is active --- identified 42 episodes with $V_{node}\approx0$ in 28 of the 72 sessions. Twenty-one of these episodes occurred at operating currents of 374--585 A on eight magnets from five families. The median estimate of $R_{lead}$ from these episodes is 2.288 m$\Omega$, while the median from persistent-mode tails in another subset is 2.542 m$\Omega$. Agreement in order of magnitude supports use of the shorted-node signature at operating current but does not transfer the Avanto PCS characteristic to these magnets.

The selected episodes are not retrapping events. Zero node voltage by itself does not determine PCS state: $V_{node}\approx0$ is observed both with a superconducting branch shorting the node ($R\to0$, possibly with a large branch current) and with a resistive branch carrying nearly zero current (finite $R$, but $I_{sw}\to0$, so $V_{node}=R_{sw}I_{sw}\to0$). The identified segments are periods in which power-supply current is aligned with coil current before the PCS is driven resistive. Both cases occur in sequence within these segments: before the heater is turned on, the node is shorted by the superconducting branch; after the PCS opens, the resistive-branch current is near zero because the currents have already been matched. Thus, the heater-on transition appears inside the segment, opening the PCS changes the coil current negligibly, and the subsequent current change begins later. The same stationarity condition is required before trapping current in persistent mode.

No evidence of PCS return to the superconducting state due to loss of a resistive operating point was found in the field corpus. This negative result shows only that such an event did not occur among the observed operations. It is not a test of the criterion in Section 5.4, because neither $R(V,h)$ nor the complete $R_{path}$ was measured for the field PCS devices.

\FloatBarrier
\subsection{Relaxation before current capture into persistent mode}
After the ramp is complete, the power-supply current is close to its final target value, but this does not yet imply $I_{psu}=I_{coil}$. While the PCS heater remains on, residual current in the resistive branch continues to redistribute into the coil. As shown in Section 4.3, this redistribution obeys

\[
\frac{dI_{sw}}{dt}+\frac{R_{sw}}{L}I_{sw}
=\frac{dI_{psu}}{dt},
\]
and once the power supply has settled, its natural time scale is $\tau=L/R_{sw}$.

As redistribution proceeds, the residual PCS current transfers into the coil, $V_{node}$ decreases, and the power-supply voltage approaches the value needed only to compensate the external-path voltage drop:

\[
V_p=R_{path}I_{psu}+V_{node}
\longrightarrow R_{path}I_{psu}.
\]
In the limiting settled state,

\[
V_{node}\to0,\qquad
I_{sw}\to0,\qquad
I_{coil}\to I_{psu}.
\]
This is the pattern that should be seen in the field channels by the end of the hold. A nonzero current through a resistive PCS at this stage indicates incomplete redistribution rather than a stationary branch current at $V_{node}=0$.

In actual field records, the beginning of the final hold is more complicated than this limiting case. After overcurrent is removed, the power-supply regulator is still settling to the final current. In the sessions considered, $I_{psu}$ differs from its setpoint by about 0.2--0.6 A and returns to it over tens of seconds. In 17 of 21 sessions, $V_{node}$ changes sign during this process. With strictly constant power-supply current, a free exponential relaxation cannot change sign; it is therefore incorrect to describe the entire hold by a single function $A\exp(-t/\tau)+V_{off}$.

This behavior is conveniently illustrated by the trajectory of the minimal node model in Fig. 13a: the power supply is close to its final value but is still settling; current continues to redistribute in the coil; while $I_{coil}<I_{psu}$, node voltage is positive; when the currents cross, node voltage changes sign; only after the power supply has settled does the ordinary free-relaxation tail toward zero remain.

\begin{figure}[!htbp]
\centering
\includegraphics[width=\linewidth]{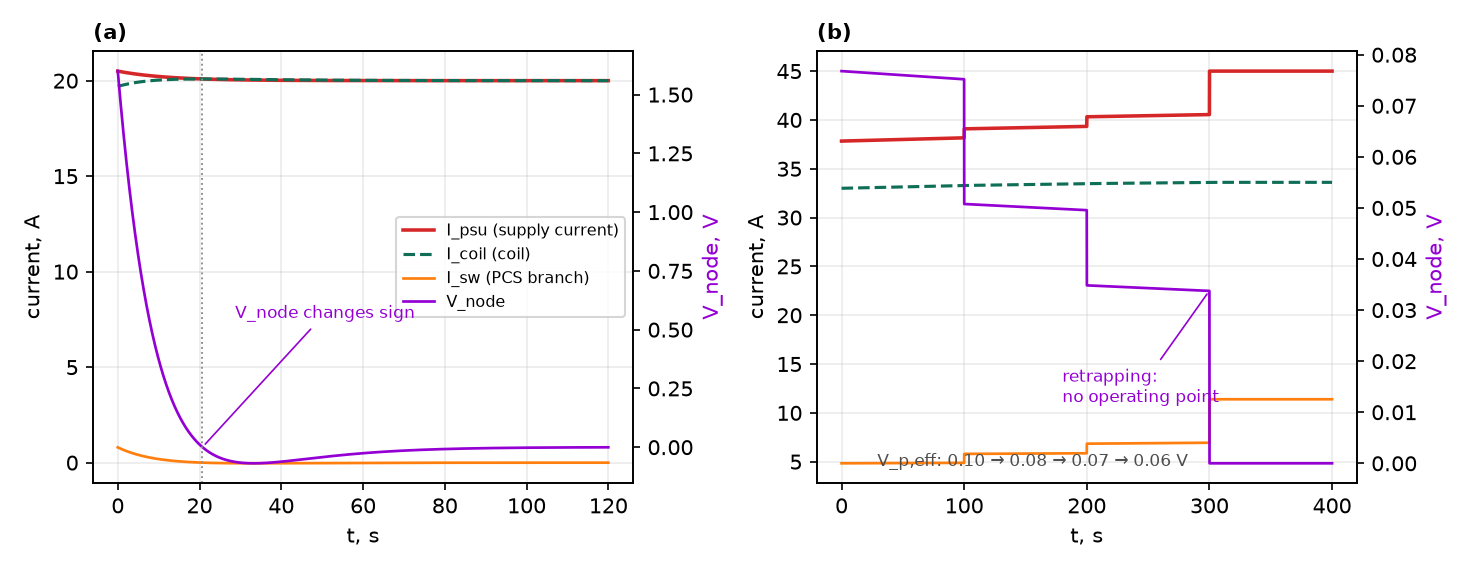}
\caption{Illustrative trajectories of the minimal coil $\parallel$ PCS node model (equations of Section 3.2 and previously identified Avanto parameters). (a) Final hold with the heater on: the power supply removes overcurrent and is still settling; when $I_{coil}$ and $I_{psu}$ cross, node voltage changes sign, so the early part of the hold is not a free exponential. (b) Approach to the self-sustain boundary with the heater off (Section 5.4): stepwise reduction of $V_{p,eff}$ increases branch current, and below the tangency the resistive operating point disappears --- retrapping occurs. The curves are intended only to visualize mechanisms derived from the node equations; they are neither an independent validation nor a fit to a particular field session. The power-supply regulator and PCS thermal dynamics are not modeled, and retrapping is instantaneous.}
\label{fig:13}
\end{figure}
For this part of the hold, the same complete mismatch equation is used, with measured $I_{psu}(t)$ treated as the system input. Using

\[
I_{sw}=\frac{V_{node}}{R_{sw}},
\qquad
L\frac{dI_{coil}}{dt}=V_{node},
\]
its integral form is

\[
I_{psu}(t)
-\frac{1}{L}\int_{t_0}^{t}V_{node}(t')\,dt'
=
\frac{V_{node}(t)}{R_{sw}}+I_{coil}(t_0).
\]
For a fixed window this is a linear least-squares problem in $1/R_{sw}$ and the unknown $I_{coil}(t_0)$. The estimator does not require the power-supply current to be constant. An estimate is accepted only if the span of $V_{node}$ is at least 0.05 V, the branch-current contribution is at least five times the residual, and the result is stable when a linear $V_{node}$ drift is added.

Applying the balance to two portions of the same hold shows that the effective PCS resistance does not remain constant along the hold. In all 11 sessions where both portions can be determined, the resistance on the higher-voltage portion is 4.5--10.7 $\Omega$, exceeding the resistance of the low-voltage tail $|V_{node}|\le0.15$ V, which is 2.5--5.7 $\Omega$. The direction of this change is qualitatively consistent with the electrothermal description of Section 5.5 for the heater-on state: at higher voltage, Joule heating $V_{node}^2/R_{sw}$ adds to the heater input and increases the effective resistance; as $V_{node}$ approaches zero, the low-voltage nearly ohmic regime remains. For Avanto magnets in the archival corpus, the low-voltage values are 1.8--3.8 $\Omega$, partially overlapping the 1.86--2.19 $\Omega$ range directly measured on the PCS studied here.

For capture error, the end of this relaxation is what matters. At the moment the heater-off command is issued, the residual current in the resistive branch is estimated as

\[
I_{sw}=\frac{V_{node}}{R_{low}},
\]
where $V_{node}$ is the median over the final six seconds before the command, and $R_{low}$ is the resistance of the low-voltage tail determined for that same session by the balance described above. No PCS coefficients from the Avanto studied in Section 5 are used here. $V_{node}$ is reconstructed by signed subtraction of $R_{lead}I_{psu}$; $R_{lead}$ for each session is determined from the subsequent persistent-mode tail, while the inductance of each magnet is determined in advance from ramps in that magnet's own logs.

The corpus includes 21 sessions with operating current, recorded heater state, and previously determined inductance. In 17 of them, $R_{low}$ could be determined directly; its median is 3.5 $\Omega$, with a range of 1.8--5.7 $\Omega$. The median residual PCS current immediately before the heater-off command is 5.0 mA, or 10.7 ppm of the coil current; the maximum among these 17 sessions is 15 mA, or 30 ppm.

This value is not a directly measured final error in persistent current. Transition of the PCS to the superconducting state after heat is removed is not instantaneous: while the cooling branch remains resistive, the existing $V_{node}$ continues to change the coil current and part of the remaining $I_{sw}$ can still transfer into the coil. Therefore, the $I_{sw}$ calculated at the command time is a conservative upper bound on the final mismatch of the captured current. No independent magnetic-field measurement is present in these logs; current ppm is used only as a proxy for field ppm under the approximately linear relationship between field and current, and no formal field tolerance is established here.

In four sessions, $R_{low}$ cannot be determined directly. In two of them, the signal before heater switch-off is small: $V_{node}$ is approximately $-$0.1 mV and $-$13 mV, corresponding to a small residual --- of order 10 ppm or less over the observed $R_{low}$ range. In the other two, the low-voltage tail is shorter than 10 s and the residual voltage is appreciable: $-$98 and $-$171 mV. For $R_{low}=1.8-5.7\ \Omega$, these correspond to 37--115 and 59--182 ppm, respectively. The latter episode is the worst case in the corpus: before heater switch-off, $V_{node}$ did not decay monotonically but oscillated with a period of about 30 s, and the command was issued in the middle of an oscillation.

It is useful to characterize the adequacy of the final hold by the time elapsed after the power supply itself has settled to the final current. For the low-voltage tail define

\[
\tau_{low}=\frac{L}{R_{low}}.
\]
After $I_{psu}$ has settled, the dynamics approach free redistribution, and $\tau_{low}$ provides the natural time scale over which the residual PCS current transfers into the coil. Across the 17 sessions, the time from power-supply settling to heater switch-off has a median of $3.6\tau_{low}$ and a minimum of $2.4\tau_{low}$. For constant resistance,

\[
\exp(-3.6)\approx0.03,
\]
so after such an interval about 3\% of the PCS current present at the beginning of the low-voltage tail remains. This 3\% cannot be compared directly with 10.7 ppm: the former is normalized to the initial PCS branch current, while the latter is normalized to the full coil current.

For the magnet with $L=111.5$ H, the actual hold is 177--180 s, compared with 28--70 s on the other systems, but expressed in units of $\tau_{low}$ the range is comparable, 3.0--3.8. This is consistent with scaling the required time with $L/R_{low}$, although it does not by itself show that the routine service procedure was specified in terms of this time constant.

Thus, final current capture has two sequential conditions. First, the power supply itself must settle; then, the residual current in the resistive PCS branch must have enough time to transfer into the coil. The electrical signature of completion of the second process is a stable approach of $V_{node}$ to zero. A single fixed hold duration is therefore insufficient: both the shape of the observed relaxation and the number of elapsed $\tau_{low}$ time constants are relevant.

\FloatBarrier
\subsection{Fast noninductive component of power-supply current at operating current}
Sections 6.1 and 6.2 tested consequences of the circuit relevant to the operating procedure. This section tests a consequence that follows directly from the topology and contains none of the coefficients of the $R(V,h)$ characteristic.

The terminal current is the sum of currents in the two parallel branches, $I_{psu}=I_{coil}+I_{sw}$, while the coil current is limited by its inductance: over an interval $\Delta t$, it cannot change by more than $L^{-1}\int V_{node}dt$. Therefore, if the observed change in $I_{psu}$ substantially exceeds this amount, it is not the coil component that changed. In engineering terms, the coil physically cannot change its current by the observed amount --- the measured voltage is several times smaller than required --- so the rapidly changing portion of the measured current must belong to another, parallel branch. The inductance of each magnet is determined in advance from ramps in its own logs (Section 3.5), $V_{node}$ is reconstructed by the signed subtraction of Section 3.2, and the correction $R_{lead}\Delta I_{psu}$ across a fast step is about 2 mV compared with hundreds of millivolts in the step itself, so it does not affect the conclusion.

A fast change in $I_{psu}$ alone is not sufficient evidence. Such changes also occur when the PCS itself switches, when regulation transfers between current and voltage loops at ramp boundaries, and during an emergency discharge. Events were therefore selected using recorded power-supply mode channels rather than by visual appearance of the curve. The selection was applied to all 56 field sessions in the archive; the heater-state gate uses the CAN trace recorded in 49 of them. Events were accepted only when the heater state did not change within $\pm15$ s, the power-supply operating mode remained constant, the regulator did not switch between current and voltage regulation during the response itself, and current before and after the step drifted at a rate consistent with $V_{node}/L$, i.e. behaved like coil current outside the step. Events at ramp boundaries and during mode changes were excluded entirely. In one session, the recorded CAN structure is empty; the stated heater-state gate cannot be verified for that event, so it is excluded from the statistical corpus and shown separately as an additional event (sheet B7 of Appendix B).

Across the 56 field sessions, an independent re-analysis retained 19 electrically clean events. After removing the event with unverifiable heater-state gates, the final statistical corpus consists of 18 events in 17 sessions on 12 systems from four families; 15 occurred at operating currents of 450--573 A. In all 18 events, the observed rate of change of power-supply current would have required a coil voltage 5--46 times greater than that actually measured at the node, with a median factor of 11.

This is the primary result of the section: it is local and does not depend on the time interval over which integration is performed. The corpus is summarized in Fig. 14. Panel (a) shows the factor by which the observed current-change rate exceeds the maximum possible coil rate at the measured $V_{node}$, thereby demonstrating that the fast component cannot be coil current. Panel (b) addresses the next question --- the magnitude of that component --- over the full current range.

\begin{figure}[!htbp]
\centering
\includegraphics[width=\linewidth]{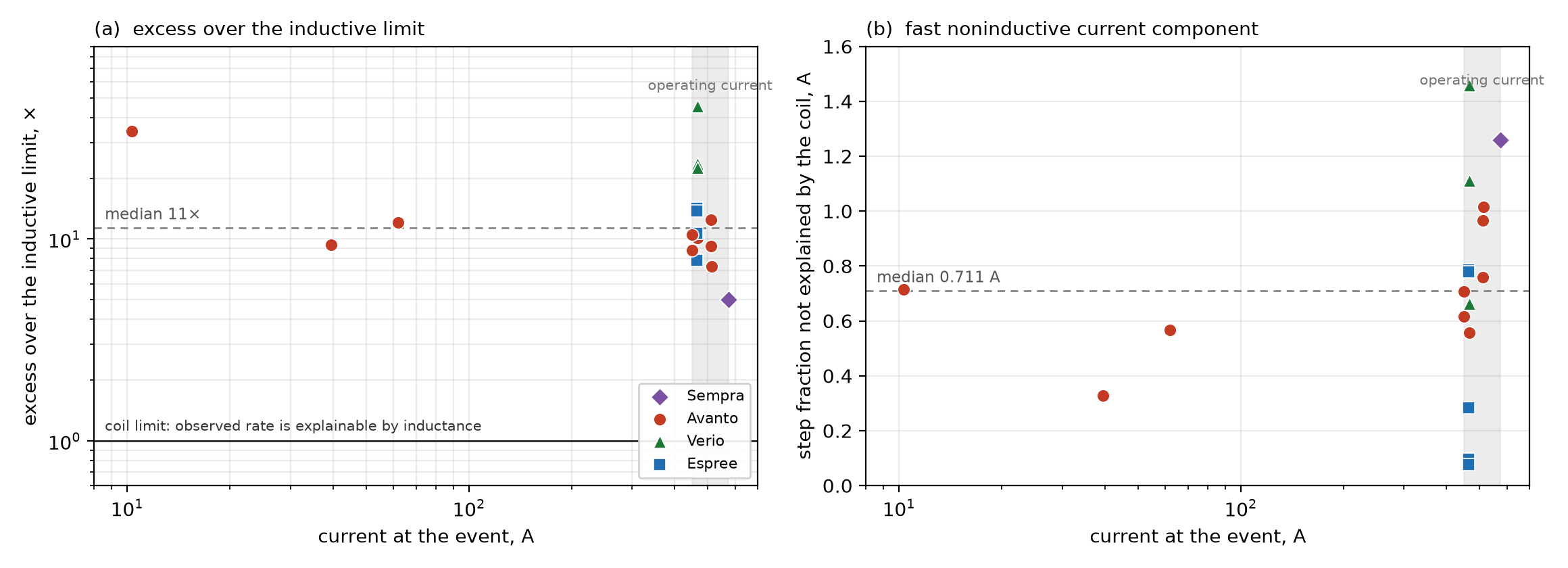}
\caption{Summary of the statistical corpus for the fast noninductive component: 18 events in 17 sessions on 12 magnets from four families (the additional event with an empty CAN structure is excluded from the statistics; see Appendix A). (a) Excess over the inductive limit: the factor by which the observed power-supply current-change rate exceeds the maximum possible coil rate at the measured node voltage, $M=L|dI/dt|_{\max}/\max|V_{node}|$. The line $M=1$ is the physical boundary: below it, the change could still belong to the coil; above it, it cannot. All 18 events lie above the boundary: median 11$\times$, range 5--46$\times$. (b) Fast noninductive current component: the portion of the step not explained by the coil using the conservative window; median 0.711 A, range 0.075--1.459 A. The gray band in both panels marks operating current 450--573 A: 15 of 18 events occurred in this range rather than in low-current diagnostic maneuvers. Numerical values for every event are given in the Appendix A table; enlarged plots are collected in the Appendix B atlas.}
\label{fig:14}
\end{figure}
Integral decomposition of the step provides a second, more intuitive formulation. The portion attributed to the coil depends on the chosen time window because $\Delta I_{coil}$ accumulates as an integral and grows with the window. A longer window --- extending to the settling of node voltage --- was therefore used. This maximizes the current change attributed to the coil and thus yields a conservative estimate of the residual. Even with this choice, the median fraction of the step not explained by coil-current change is 69\%, with a minimum of 25\%. The systematic effect of the window definition itself was assessed by comparison with a shorter selection rule and is 0.15 A at the median and 0.50 A in the worst case.

Figure 15 shows two cases on magnets from the same family at currents differing by a factor of fifty. In the first, only the voltage setpoint changes while the heater state and power-supply mode remain unchanged; this is the same event shown in Fig. 1c. In the second, at an operating current of 511 A, an unloading command reduces both setpoints; the power-supply current decreases by 1.339 A, whereas the coil can change by only 0.372 A under the measured node voltage over the same interval, leaving 0.967 A unexplained by inductance. A third event from the same corpus is shown in Fig. 1d: the voltage setpoint increases at 465 A on the same magnet as the low-current maneuver; the step is 0.795 A, the coil can account for 0.240 A, and the residual is 0.556 A.

\begin{figure}[!htbp]
\centering
\includegraphics[width=\linewidth]{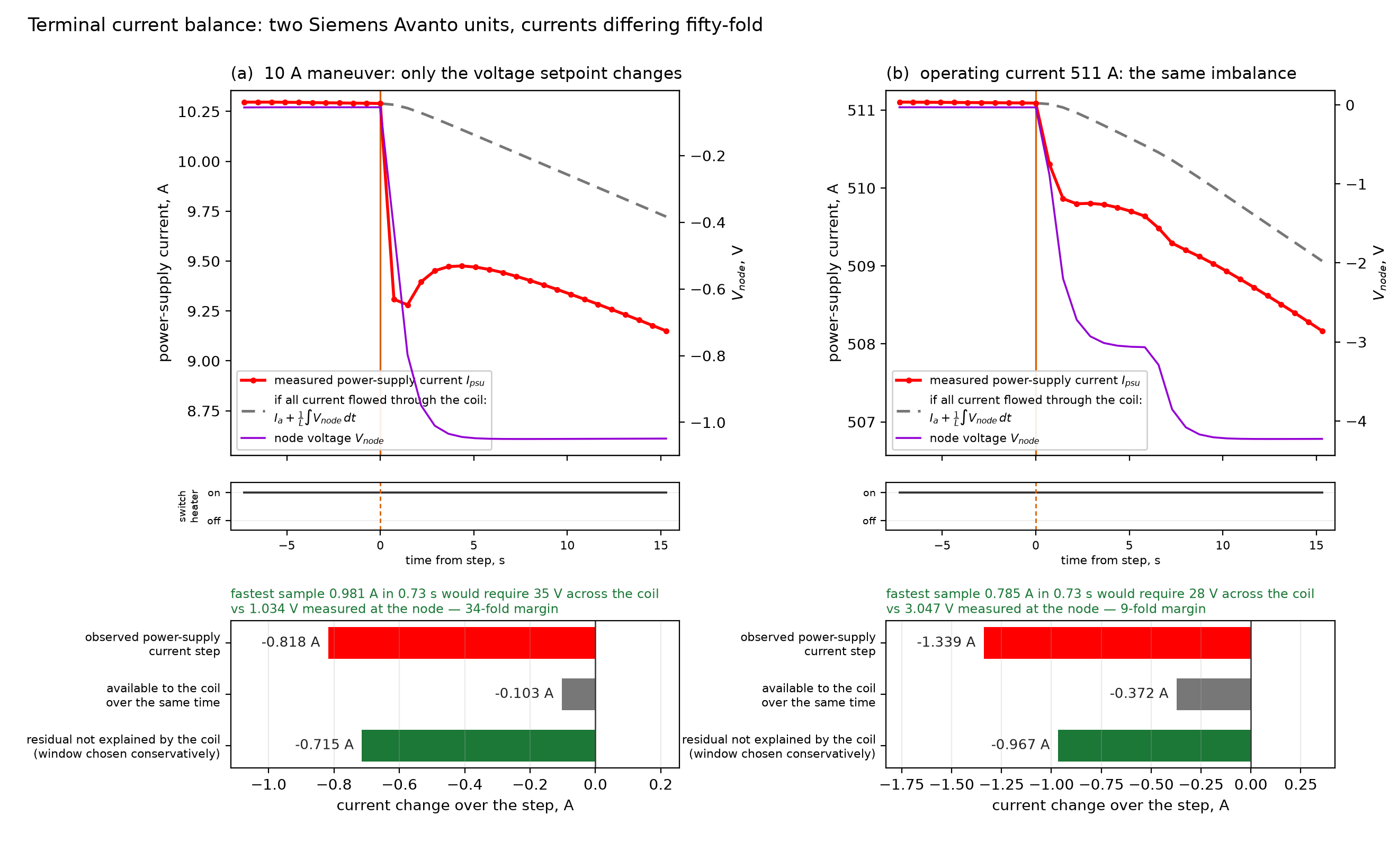}
\caption{Terminal current balance at fixed PCS state for two Siemens MAGNETOM Avanto 1.5 T magnets from the Section 3.4 archive. Upper panels: red curve, measured power-supply current; purple curve, reconstructed node voltage; gray dashed curve, counterfactual trajectory $I_a+L^{-1}\int V_{node}dt$, i.e. the current that would be observed if the entire power-supply current flowed through the coil. Below each panel is the recorded heater state, unchanged throughout the window. Lower panels: decomposition of the observed step into the portion available to the coil according to this integral and the residual. (a) 10 A maneuver, only the voltage setpoint changes: step $-$0.818 A, coil-available change $-$0.103 A, residual $-$0.715 A; the fastest sample, $-$0.981 A in 0.73 s, would have required 35 V across the coil compared with the measured 1.03 V at the node. (b) Operating current 511 A, unloading command: step $-$1.339 A, coil-available change $-$0.372 A, residual $-$0.967 A; the fastest sample would have required 28 V compared with the measured 3.05 V. In both cases the step window is extended until node voltage settles, maximizing the portion assigned to the coil; branch resistance is not estimated from these data.}
\label{fig:15}
\end{figure}
Under the adopted topology, the only path parallel to the coil is the resistive PCS branch, and the residual is attributed to it.

What is demonstrated here is the existence of a separate fast branch --- and only that. The $R(V,h)$ coefficients were not identified for these magnets, and the ratio $\Delta V/\Delta I$ during the transient is a differential quantity and is not identified here with the quasi-static $R_{sw}=V/I_{sw}$ of Section 5. The only test is that the separate fast component of power-supply current predicted by the topology is indeed observed, including at the operating currents of installed magnets. A full audit of the corpus is provided in the appendices: selection criteria, corpus provenance, and event tables in Appendix A; an atlas of the 18 statistical events and additional event B7 in Appendix B; and examples of excluded events in Appendix C.

\FloatBarrier
\section{Discussion}
\FloatBarrier
\subsection{Physical meaning of the two resistive regimes}
The PCS has two resistive-branch characteristics: a nearly ohmic one at $h=1$ and a self-sustaining power-law one at $h=0$. The heater initiates the transition and changes the thermal balance, but it does not uniquely determine the PCS state. The unified law

\[
R=C\left(P_hh+\frac{V^2}{R}\right)^a
\]
supports an electrothermal interpretation without determining the temperature or geometry of the normal zone; $P_h$ remains an effective parameter.

The resistance of the resistive branch is not fixed. With the heater on, it is 1.86--2.19 $\Omega$ in the measured windows below 0.5 V, 3.64 $\Omega$ at 2.974 V, and a separate point at 3.95 V in another session gives about 11 $\Omega$. The intermediate characteristic between 0.5 and 3.9 V is not resolved by stable segments and is not interpolated (Section 8).

One experiment shows how strongly the resistive state at these voltages depends on supplied heat. On the upper plateau of the 3.0 $\to$ 0.2 V level sequence, the heater was switched off halfway through the plateau. At nearly unchanged node voltage, branch resistance was 3.64 $\Omega$ with the heater on (2.974 V) and 2.47 $\Omega$ after it was switched off (2.926 V). Heater switch-off reduces $R_{sw}$ by 32\% relative to its heater-on value; equivalently, heater-on raises the resistance by 47\% relative to heater-off.

Thus, even near 3 V the branch resistance remains substantially sensitive to additional thermal input. No evidence was seen that the branch had reached a saturated resistive regime largely independent of heater state. External measurements cannot distinguish what changes physically --- the length of the resistive region, its temperature, or both. Such a distinction would require the temperature dependence of the conductor's normal resistance, which is not known here.

The full normal resistance of the PCS studied is not determined by the present measurements. No direct estimate is available from this work; published values refer to another construction and cannot be transferred to this switch (Section 2). No dedicated measurements above 3.95 V were performed, so the location of any possible saturation of the characteristic was not tested. External access measures only the equivalent branch and cannot resolve whether it is formed by one resistive element or several parallel elements; therefore, $R_{sw}$ is used throughout as the effective resistance of the PCS resistive branch as a whole.

The same pair of points also shows how the difference between regimes narrows: at 0.4 V, the two resistances differ by about a factor of twelve, whereas at a node voltage near 2.95 V they differ by only about a factor of 1.5. The mechanism follows from the unified balance of Section 5.5: as voltage increases, internal dissipation $V^2/R$ dominates the heater contribution, so changing $h$ shifts the operating point less strongly. Quantitatively, however, the relation with the same coefficients predicts a difference of about a factor of two here rather than the measured factor of 1.5, and both points lie outside its fitting window. The observation should therefore be treated as qualitative support for the mechanism, not as an independent validation of the numerical model.

\FloatBarrier
\subsection{Observable self-sustain boundary in the installed system}
The self-sustain boundary observable through external electrical variables is determined by the intersection of the identified PCS characteristic and a load line that depends on $R_{path}$ and $I_{coil}$. Its terminal position therefore belongs to the installed system ``PCS + external path + coil current,'' not to the PCS alone. In instrument studies of retrapping, the measurement circuit plays the role of the load \cite{r3,r4,r5}; in an installed magnet, this role is played by the cable and current-lead resistance together with the magnet current itself. This does not exclude the existence of an intrinsic thermal boundary of the normal zone under other conditions; such an internal boundary is not separately identified in this work.

\FloatBarrier
\subsection{Scale of the external-path correction at nominal current}
The dedicated maneuvers were performed at currents up to 42 A, whereas the operating current of the magnet is about 500 A. This matters for the observable self-sustain boundary because part of the terminal voltage is lost across the resistance of the external path. With $R_{path}=4.802$ m$\Omega$, this drop is about 0.20 V at 42 A and 2.40 V at 500 A.

In Section 5.4, the boundary was expressed through the effective voltage

\[
V_{p,eff}=|V_p|-R_{path}|I_{coil}|,
\]
and for the PCS studied $V_{p,crit}=0.0619$ V. If the PCS characteristic and $R_{path}$ measured at low current remain valid at 500 A, the same boundary would correspond to an external terminal voltage

\[
|V_p|\approx R_{path}|I_{coil}|+V_{p,crit}
\approx2.40+0.0619\approx2.46\ \text{V}.
\]
It is useful to write the three voltage levels explicitly:

\[
|V_p|\approx2.46\ \text{V}
\quad\rightarrow\quad
V_{p,eff}\approx0.0619\ \text{V}
\quad\rightarrow\quad
V_{node,*}\approx0.0189\ \text{V}.
\]
The 2.46 V level is the voltage that the power supply would have to produce at the external terminals; 0.0619 V is what remains after subtracting the cable and current-lead drop and determines the position of the load line; 0.0189 V is the actual voltage directly across the PCS at tangency. This scenario applies to heater-off intervals. At $h=1$, the branch is ohmic, self-sustain is not required, and the boundary does not apply.

The value 2.46 V is not a new PCS boundary and was not measured at operating current; it is only a conversion of the low-current boundary to the external terminals for a hypothetical current of 500 A. Testing this extrapolation requires dedicated maneuvers at operating current. The field events confirm current redistribution at operating current (Section 6.3), but do not test the numerical extrapolation of the boundary.

\FloatBarrier
\subsection{Implications for the hold, procedure, and ramp model}
The natural scale of the final hold is $\tau_{low}=L/R_{low}$. For a particular magnet, $R_{low}$ can be determined from previous final holds --- exactly as it was obtained, without dedicated maneuvers, in 17 of the 21 field sessions in Section 6.2 --- or by dedicated identification, and then used as a parameter of the procedure for that specific magnet. A preliminary estimate by magnet class is possible, but the spread of low-voltage values in the corpus (1.8--5.7 $\Omega$) makes it only an initial approximation.

The number of elapsed $\tau_{low}$ should be considered together with the observed $V_{node}$: heater switch-off is appropriate only after the power supply has settled and $V_{node}$ is stably approaching zero. The field medians --- a margin of $3.6\tau_{low}$ after power-supply settling and a residual of 10.7 ppm at the heater-off command (a conservative upper bound on the final mismatch; Section 6.2) --- describe the observed procedures but do not define a standard. The case with oscillating $V_{node}$ ($-$171 mV at the node at heater switch-off, corresponding to 59--182 ppm over the corpus range of $R_{low}$) shows that in the absence of a stable decay, the time-to-$\tau$ ratio alone is insufficient.

A ramp model must include the input $h$ and the discrete PCS state, while external identification must use the signed voltage $V_{node}=V_l-R_{lead}I_{psu}$ and the current-separation methods of Section 4. Such a model is needed primarily so that ramp procedures can be evaluated in simulation rather than by maneuvers on an installed clinical system. Construction and validation of that full model are the subject of separate work.

\FloatBarrier
\subsection{Methodological implications of external identification}
External PCS identification is most sensitive to determination of node voltage. The full power-supply current flows through the current leads, so the correct correction is

\[
V_{node}=V_l-R_{lead}I_{psu}.
\]
If $I_{coil}$ is used instead of $I_{psu}$, the reconstructed voltage retains the drop $R_{lead}I_{sw}$, and the calculated resistance becomes $R_{sw}+R_{lead}$. For the characteristic studied here, the relative distortion is about 0.5\% at the upper end of the range and reaches 21\% at the lower end. Such a large percentage despite milliohm current leads is explained by the branch itself: in the low-voltage part of the power-law characteristic, $R_{sw}$ falls to about 12 m$\Omega$ at 0.07 V, so several milliohms of $R_{lead}$ are no longer a small correction. The definition of the derived quantity $V_{node}$ must therefore be part of the definition of the reported $R(V)$ law itself.

A second difficulty is current separation. With the heater on, branch current was 0.1--0.2 A against a coil current of 20--30 A, so reconstructing it as the difference $I_{psu}-I_{coil}$ is poorly conditioned: a small error in the assumed initial coil current is comparable with the quantity being estimated. In the self-heated regime, branch current is 2--3 A, so the same error has a much smaller relative effect. This explains why ordinary integral reconstruction is particularly unreliable in the low-voltage heater-on regime.

The methods of Section 4 remove this dependence in different ways: they use continuity of coil current at the falling edge, two zero-voltage windows, or the integral of the remaining decay. Low-coil-current maneuvers additionally make branch current comparable with the measured total current. In this way, the characteristic of an installed PCS can be obtained from external electrical channels without access to the internal node and without assuming an unknown initial coil current. The applicability of the result is then determined not only by instrument accuracy, but also by whether the chosen maneuver creates a sufficiently well-conditioned measurement situation.

\FloatBarrier
\section{Limitations}
The $R(V,h)$ laws, electrothermal parameters, and observable terminal self-sustain boundary were obtained on one Siemens MAGNETOM Avanto PCS in dedicated maneuvers at currents up to 42 A. No dedicated experiments were performed to test these quantitative relations at the operating current of approximately 500 A, so their persistence in that range has not been established and the estimate in Section 7.3 remains a scenario. They also cannot be transferred to other PCS devices without separate identification.

Field logs at 374--585 A are used only to test general consequences of the electrical topology and relaxation using parameters determined for each individual session. This limitation remains fully in force after Section 6.3: that section independently observes only the existence of the fast noninductive current component predicted by the topology --- at currents up to 573 A and without using any coefficient of the PCS characteristic --- but does not transfer any of its numerical dependencies. The transient ratio $\Delta V/\Delta I$, which could be calculated from the same events, is a differential quantity; its relationship to the quasi-static $R_{sw}=V/I_{sw}$ of Section 5 was not separately investigated, and the ratio is not used in this work.

Attribution of the residual step to the PCS branch is itself a topological assumption. If the magnet contains another path in parallel with the coil that cannot be resolved by external measurement, its current would be indistinguishable from PCS current in these data; the existence of the fast noninductive component does not depend on this assumption. The size of the fraction attributed to the parallel path depends on the selected time window. Compared with a shorter selection rule, the discrepancy is 0.15 A at the median and 0.50 A in the worst case for steps of about one ampere. The longer window, which maximizes the coil contribution, was therefore adopted.

The derived tangency criterion applies to the terminal boundary of the resistive operating point within the measured external circuit. It does not determine a possible intrinsic thermal PCS boundary in the limit $R_{path}\to0$. The power law of the $h=0$ regime was measured only at finite $|V_{node}|$ (0.025--2.4 V) and should not be extrapolated to $V_{node}=0$, where the resistive description of the branch ceases to apply. The upper boundary of the window is also a limit of applicability: preservation of the power law at node voltages above 2.4 V, characteristic of field ramps, has not been established.

Validation of the boundary is also limited to this PCS: 17 segments are inter-session and another seven are internal. Dynamics were reproduced on another session with fixed coefficients, but the selected plateau had previously been used in choosing the model configuration; this is not a fully independent held-out test. The quantity $h$ is the heater state reported over CAN by the magnet-supervision module, not a measurement of heater current, temperature, or power. The available data do not distinguish whether that report reflects actual power dissipation in the heater or only an accepted command. The parameter $P_h$ is effective. The unified balance is fitted to the same 18 points used to compare the two regimes, and a complete state machine for an arbitrary ramp is not validated.

The ohmic regime is confirmed only up to 0.5 V. A separate point at 3.95 V gives about 11 $\Omega$, but the intermediate characteristic is unknown. Settling times shorter than 109 s and electrical initiation of a cold PCS without the heater are also not determined. In field data, $R_{low}$ and residual current are derived estimates; no complete uncertainty budget, parameter confidence intervals, or independent field measurement is available. The balance estimate of $R_{low}$ was obtained in 17 of 21 sessions with operating current and recorded heater state. For those sessions, the median residual current at the heater-off command is 10.7 ppm --- a conservative upper bound on the final capture mismatch --- and the median margin after power-supply settling is $3.6\tau_{low}$. These values describe the observed procedures and do not establish a normative requirement. When $V_{node}$ is unsettled or oscillatory, the time-to-$\tau_{low}$ ratio alone is insufficient, and the absence of identified retrapping on other magnets does not validate the numerical Avanto boundary.

\FloatBarrier
\section{Conclusion}
External measurements revealed two resistive regimes of the Siemens MAGNETOM Avanto PCS: a nearly ohmic regime with the heater on and a self-sustaining power-law regime after heater switch-off. Both proved to be special cases of one effective electrothermal balance,

\[
R_{sw}
=C\left(P_hh+\frac{V_{node}^2}{R_{sw}}\right)^a,
\]
in which resistance is determined by the combined thermal contribution of the heater and Joule self-heating. A single form described the measured steady resistances over a 690-fold range. Heater state is therefore a control input to the thermal balance, but not an unambiguous label of PCS state.

The $R(V,h)$ characteristic links current separation and hold duration to the final-relaxation scale $\tau_{low}=L/R_{low}$, set by the low-voltage effective branch resistance, and links the observable terminal retrapping boundary to the external-circuit load line. The inter-session estimate of 0.0619 V lies between the observed retrapping at 0.06 V and persistence at 0.07 V, and the criterion retrospectively separated the outcomes of 24 archived segments.

For the dynamic model, the input $h$ is explicitly required: the fixed pair of characteristics reduced current RMSE from 1.449 to 0.154 A, whereas models without $h$ produced no improvement. Field logs from other magnets are consistent with general consequences of the circuit. The residual branch current at the heater-off command --- a conservative upper bound on the final capture error --- has a median of 10.7 ppm, with a median margin after power-supply settling of about 3.6 low-voltage relaxation time constants. The effective resistance of the heater-on branch increases with node voltage along a single hold, providing a field manifestation of the same self-heating mechanism.

In addition, 18 selected field events on 12 magnets contained a fast component of power-supply current that cannot be explained by coil inductance: the voltage that would have been required for the coil to produce the observed current change was 5--46 times greater than that measured at the node. The quantitative PCS laws reported above pertain to the PCS studied in dedicated maneuvers at currents up to 42 A; the field-event result, by contrast, establishes the fast noninductive component at operating currents up to 573 A. Together, these results show that external current and voltage measurements are sufficient to describe an installed PCS not as a binary switch, but as a hysteretic branch of a dynamic magnet model.

\FloatBarrier

\FloatBarrier
\section*{Appendix A. Event Selection and Corpus Summary}
\addcontentsline{toc}{section}{Appendix A. Event Selection and Corpus Summary}
\setcounter{figure}{0}
\renewcommand{\thefigure}{A\arabic{figure}}
Section 6.3 is based on a corpus of fast noninductive events. This appendix gives the complete selection criteria, corpus provenance, and tables. Each accepted event is shown on a separate sheet in Appendix B; event numbers in the table match the sheet numbers.

Selection was performed using recorded power-supply operating-mode channels, not by visual appearance of the curve. Across 56 field sessions, a detector identified fast changes in power-supply current; each candidate was accepted only when all five conditions were satisfied simultaneously: the heater state did not change within $\pm15$ s; the power-supply operating mode remained constant within $\pm15$ s; the regulator did not switch between current and voltage regulation during the response itself, with $\pm2$ s margins; current before and after the step drifted at a rate consistent with $V_{node}/L$ within 0.05 A/s, i.e. behaved as coil current outside the step; and current did not enter a discharge within the window. The heater-state gate uses the CAN trace recorded in 49 of the 56 field sessions.

Candidates violating these conditions form the exclusion classes: A --- switching of the PCS itself; D --- ramp boundary or regulation transfer; C --- current collapse or change of power-supply mode (examples are given in Appendix C). Accepted events are divided into class B1, in which only the voltage setpoint changes, and class B2, in which the command changes the current setpoint as well. Among the 19 electrically clean events including additional event B7, there is one B1 event and eighteen B2 events; the statistical corpus contains one B1 and seventeen B2 events.

\textbf{Corpus provenance.} The initial review produced 20 B1+B2 candidates in 19 sessions. Before inclusion in the paper, the corpus was recalculated using a second, independently written code path that takes only the candidate list from the initial review and derives all quantities again using the conventions of the paper; 19 of the 20 events passed the gates. The excluded candidate is an event whose response window extends into a CV$\leftrightarrow$CC regulation transfer 5.8 s after the step; it is shown as a negative control in Fig. C2.

At the same time, the margin metric itself was changed. The initial review compared the voltage that would have been required from the coil with $V_{node}$ near the middle of the transition (median margin 31$\times$, range 8--84$\times$), whereas the paper convention uses the largest $|V_{node}|$ within the event window --- a conservative choice that gives, for the same 19 events, a median margin of 12$\times$ and a range of 5--46$\times$. A second selection stage is methodological: for event B7, the CAN structure of the record is empty, so the stated heater-state gates cannot be verified. Electrically the event is clean, but it is excluded from the statistical corpus and reported only as an additional event.

The statistical corpus therefore contains 18 events in 17 sessions on 12 magnets: median margin 11$\times$, median residual 0.711 A, with the same ranges stated above. Only values based on this convention are used throughout the paper; the systematic effect of time-window selection is assessed in Section 6.3.

Magnets are labeled M01-M13 (M11 appears only in additional event B7). The mapping between these labels and actual installations is not disclosed under the conditions governing access to the field data. Inductances were determined in advance from ramps in each magnet's own logs (Section 3.5):

\begin{center}
\begin{tabular}{llrllr}
\toprule
Magnet & Family & $L$, H & Magnet & Family & $L$, H \\
\midrule
M01 & Avanto & 26.0 & M08 & Espree & 36.76 \\
M02 & Avanto & 26.3 & M09 & Espree & 36.67 \\
M03 & Avanto & 26.2 & M10 & Sempra & 13.13 \\
M04 & Avanto & 36.34 & M11 & Verio & 112.0 \\
M05 & Avanto & 25.92 & M12 & Verio & 112.21 \\
M06 & Espree & 36.74 & M13 & Verio & 111.5 \\
M07 & Espree & 36.76 &  &  &  \\
\bottomrule
\end{tabular}
\end{center}
The statistical corpus contains 18 events. Event number equals the Appendix B sheet number; B7 is an additional event outside the statistical corpus and is listed separately after the main table. $\Delta I$ is the power-supply current step; ``coil-available'' is the change in current that could be produced by the coil under the measured $V_{node}$ over the step window; $|V|_{\max}$ is the largest $|V_{node}|$ in the window; ``margin'' is the ratio of $L|dI/dt|_{\max}$ to the largest $|V_{node}|$ in the window.

\begin{center}
\begin{tabular}{lllrrrrrr}
\toprule
Sheet & Magnet & Class & $I$, A & $\Delta I$, A & Coil-available, A & Residual, A & $|V|_{\max}$, V & Margin \\
\midrule
B1 & M10 & B2 & 573.3 & $-$1.828 & $-$0.570 & $-$1.258 & 2.60 & 5$\times$ \\
B2 & M05 & B2 & 511.9 & $-$1.362 & $-$0.347 & $-$1.015 & 2.91 & 7$\times$ \\
B3 & M01 & B2 & 511.1 & $-$1.339 & $-$0.372 & $-$0.967 & 3.05 & 9$\times$ \\
B4 & M01 & B2 & 510.9 & $-$1.089 & $-$0.329 & $-$0.759 & 2.98 & 12$\times$ \\
B5 & M13 & B2 & 466.6 & $-$1.168 & $-$0.057 & $-$1.111 & 2.26 & 46$\times$ \\
B6 & M13 & B2 & 466.2 & $-$1.725 & $-$0.267 & $-$1.459 & 7.65 & 24$\times$ \\
B8 & M12 & B2 & 465.2 & $-$0.999 & $-$0.338 & $-$0.662 & 7.72 & 22$\times$ \\
B9 & M02 & B2 & 465.2 & +0.795 & +0.240 & +0.556 & 1.97 & 10$\times$ \\
B10 & M07 & B2 & 463.8 & $-$0.464 & $-$0.180 & $-$0.284 & 2.37 & 14$\times$ \\
B11 & M06 & B2 & 463.8 & $-$0.324 & $-$0.227 & $-$0.097 & 2.51 & 14$\times$ \\
B12 & M09 & B2 & 463.7 & $-$1.017 & $-$0.230 & $-$0.787 & 2.63 & 8$\times$ \\
B13 & M08 & B2 & 463.5 & $-$1.021 & $-$0.241 & $-$0.780 & 2.63 & 11$\times$ \\
B14 & M06 & B2 & 463.5 & $-$0.307 & $-$0.231 & $-$0.075 & 2.54 & 14$\times$ \\
B15 & M02 & B2 & 450.2 & +1.155 & +0.448 & +0.707 & 3.55 & 11$\times$ \\
B16 & M03 & B2 & 450.1 & +0.941 & +0.325 & +0.616 & 2.70 & 9$\times$ \\
B17 & M04 & B2 & 62.1 & $-$0.931 & $-$0.365 & $-$0.566 & 3.85 & 12$\times$ \\
B18 & M04 & B2 & 39.5 & $-$0.631 & $-$0.302 & $-$0.329 & 3.70 & 9$\times$ \\
B19 & M02 & B1 & 10.3 & $-$0.818 & $-$0.103 & $-$0.715 & 1.03 & 34$\times$ \\
\bottomrule
\end{tabular}
\end{center}
Additional event outside the statistical corpus (empty CAN structure; heater-state gates cannot be verified):

\begin{center}
\begin{tabular}{lllrrrrrr}
\toprule
Sheet & Magnet & Class & $I$, A & $\Delta I$, A & Coil-available, A & Residual, A & $|V|_{\max}$, V & Margin \\
\midrule
B7 & M11 & B2 & 465.6 & $-$1.296 & $-$0.342 & $-$0.954 & 7.81 & 21$\times$ \\
\bottomrule
\end{tabular}
\end{center}
Events B9 and B19 are the same events shown in panels (d) and (c), respectively, of Fig. 1. Events B3 and B19 are shown enlarged in Fig. 15.

\FloatBarrier
\section*{Appendix B. Atlas of Accepted Events}
\addcontentsline{toc}{section}{Appendix B. Atlas of Accepted Events}
\setcounter{figure}{0}
\renewcommand{\thefigure}{B\arabic{figure}}
Each atlas sheet has the same structure. Panel (a) shows the complete session with the event marked, demonstrating that the event occurs on a settled segment rather than at a ramp boundary. Panel (b) is an enlargement: red curve --- measured power-supply current; purple curve --- reconstructed node voltage; gray dashed curve --- the counterfactual trajectory ``if all power-supply current flowed through the coil,'' $I_a+L^{-1}\int V_{node}dt$, starting from the pre-step level. The inset lists the magnet, inductance, current, event class, decomposition of the step (observed / coil-available / residual), and margin under the convention of the paper. The step window is extended until the node voltage settles; this maximizes the portion attributed to the coil and makes the residual estimate conservative (Section 6.3). Sheet numbering matches the table in Appendix A. Sheet B7 is the additional event outside the statistical corpus (empty CAN structure; Appendix A).

\begin{figure}[!htbp]
\centering
\includegraphics[width=\linewidth]{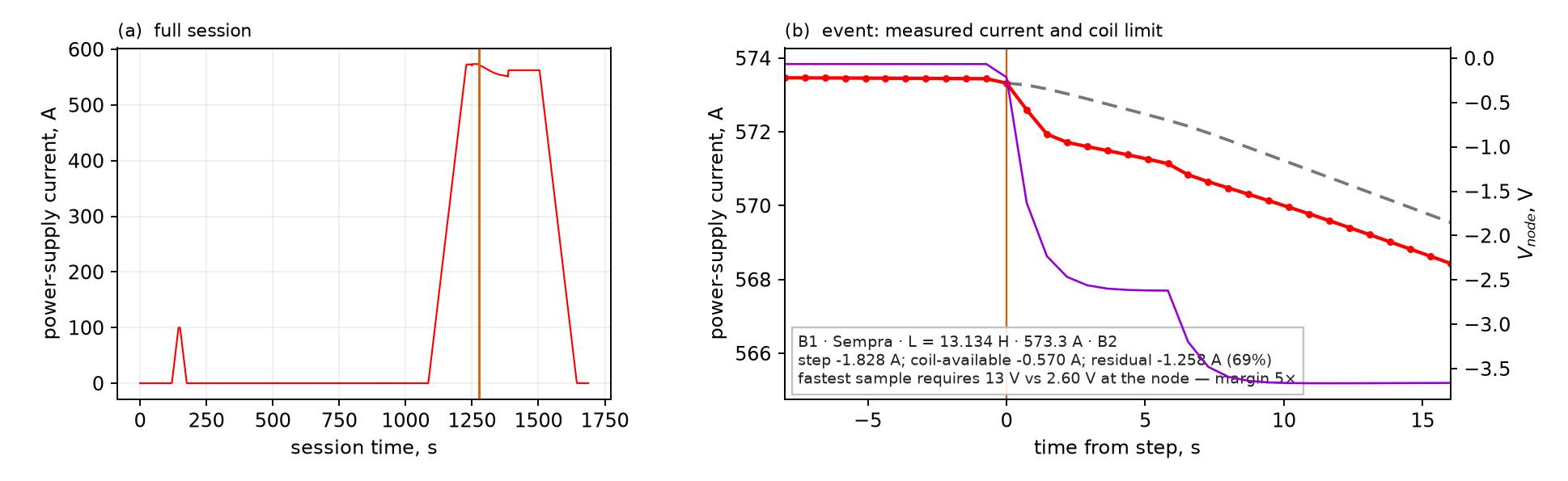}
\caption{Magnet M10 (Sempra), 573.3 A, class B2.}
\label{fig:B1}
\end{figure}
\begin{figure}[!htbp]
\centering
\includegraphics[width=\linewidth]{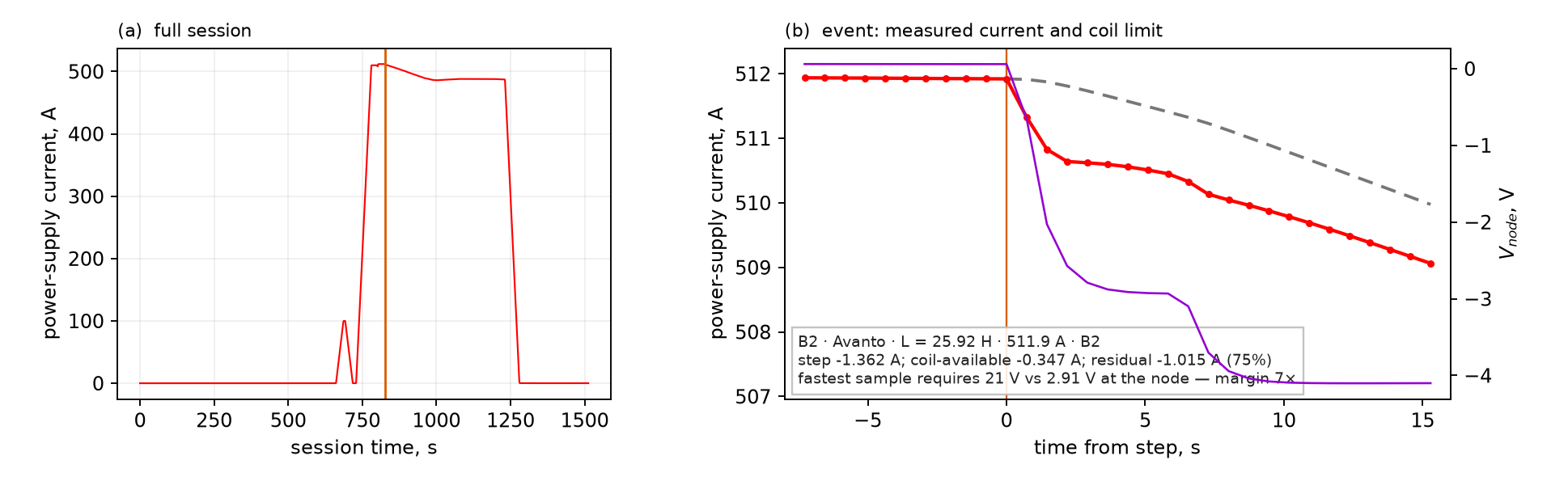}
\caption{Magnet M05 (Avanto), 511.9 A, class B2.}
\label{fig:B2}
\end{figure}
\begin{figure}[!htbp]
\centering
\includegraphics[width=\linewidth]{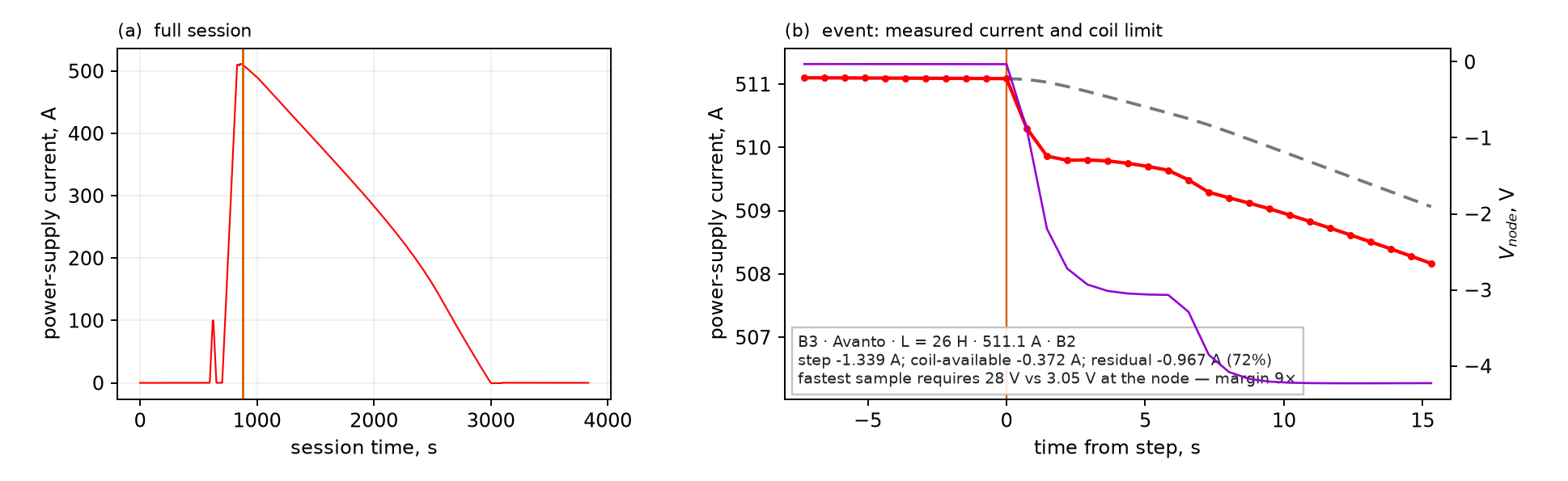}
\caption{Magnet M01 (Avanto), 511.1 A, class B2; this event is shown enlarged in Fig. 15b.}
\label{fig:B3}
\end{figure}
\begin{figure}[!htbp]
\centering
\includegraphics[width=\linewidth]{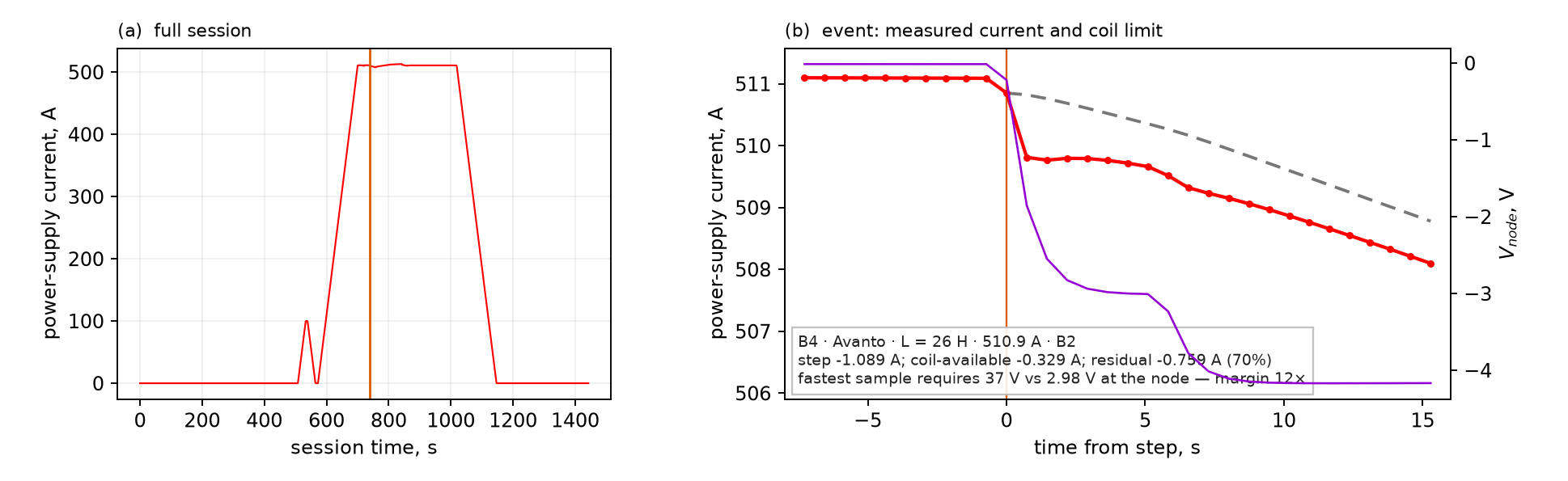}
\caption{Magnet M01 (Avanto), 510.9 A, class B2.}
\label{fig:B4}
\end{figure}
\begin{figure}[!htbp]
\centering
\includegraphics[width=\linewidth]{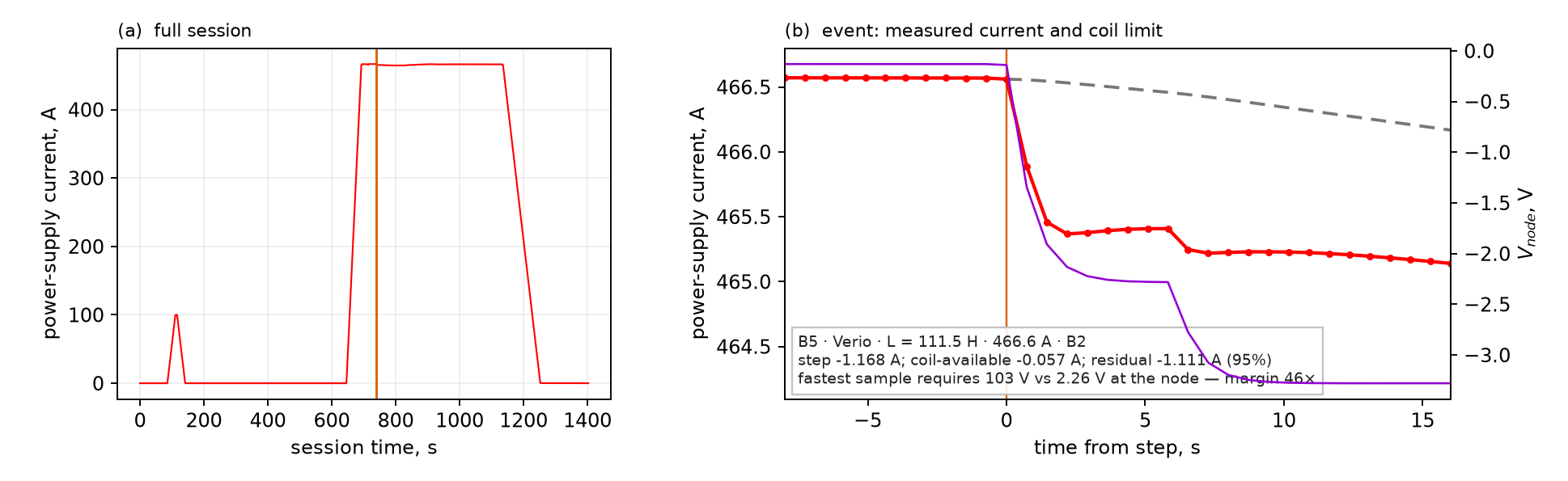}
\caption{Magnet M13 (Verio), 466.6 A, class B2; largest margin in the corpus, 46$\times$.}
\label{fig:B5}
\end{figure}
\begin{figure}[!htbp]
\centering
\includegraphics[width=\linewidth]{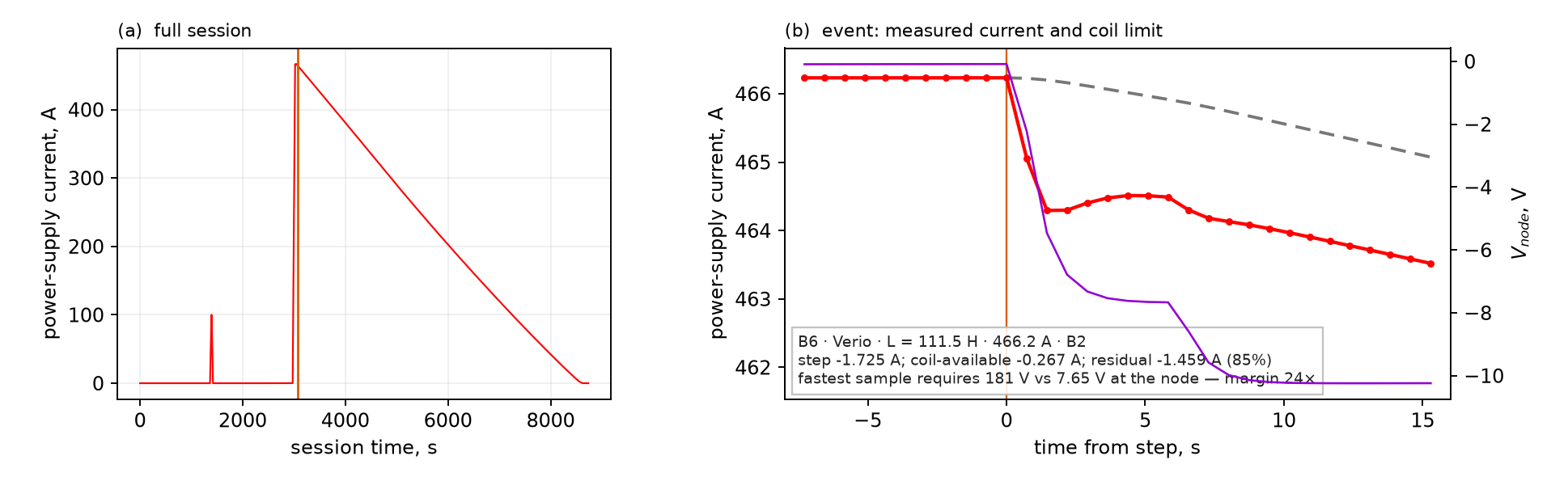}
\caption{Magnet M13 (Verio), 466.2 A, class B2; largest residual in the corpus, 1.459 A.}
\label{fig:B6}
\end{figure}
\begin{figure}[!htbp]
\centering
\includegraphics[width=\linewidth]{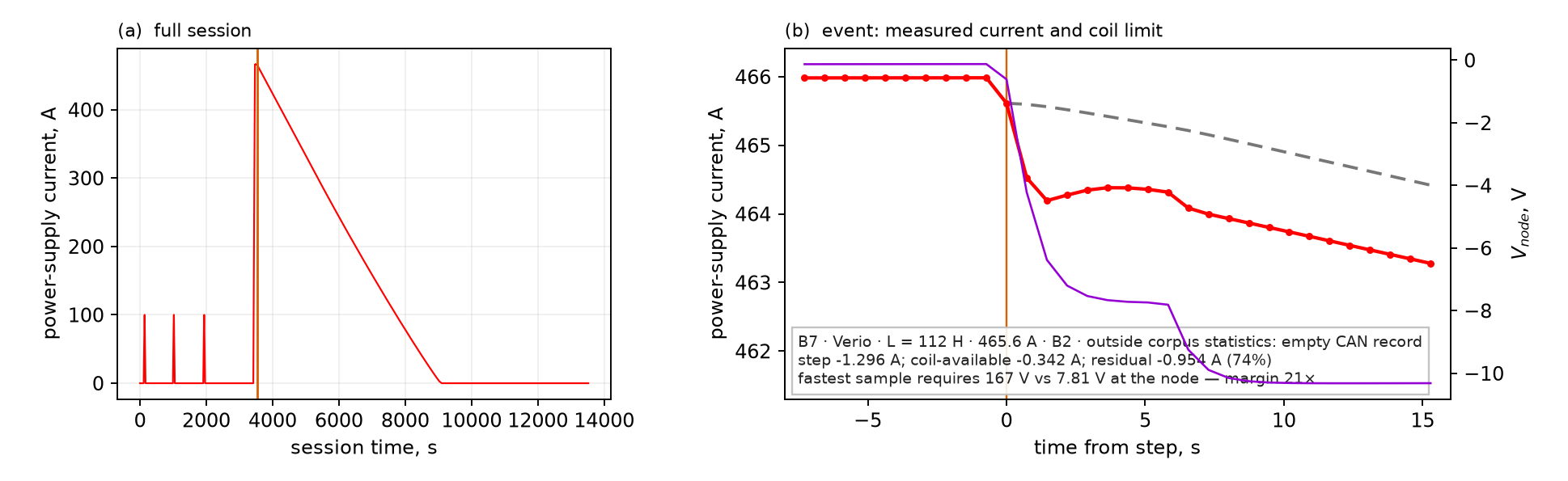}
\caption{Additional event outside the statistical corpus: magnet M11 (Verio), 465.6 A, class B2. The record contains an empty CAN structure, so the stated heater-state gates cannot be verified. All other conditions are satisfied: power-supply mode is constant, there are no CV$\leftrightarrow$CC transitions, current drift outside the step agrees with $V_{node}/L$, and the step is synchronous with the recorded unloading command. The event is not included in corpus statistics.}
\label{fig:B7}
\end{figure}
\begin{figure}[!htbp]
\centering
\includegraphics[width=\linewidth]{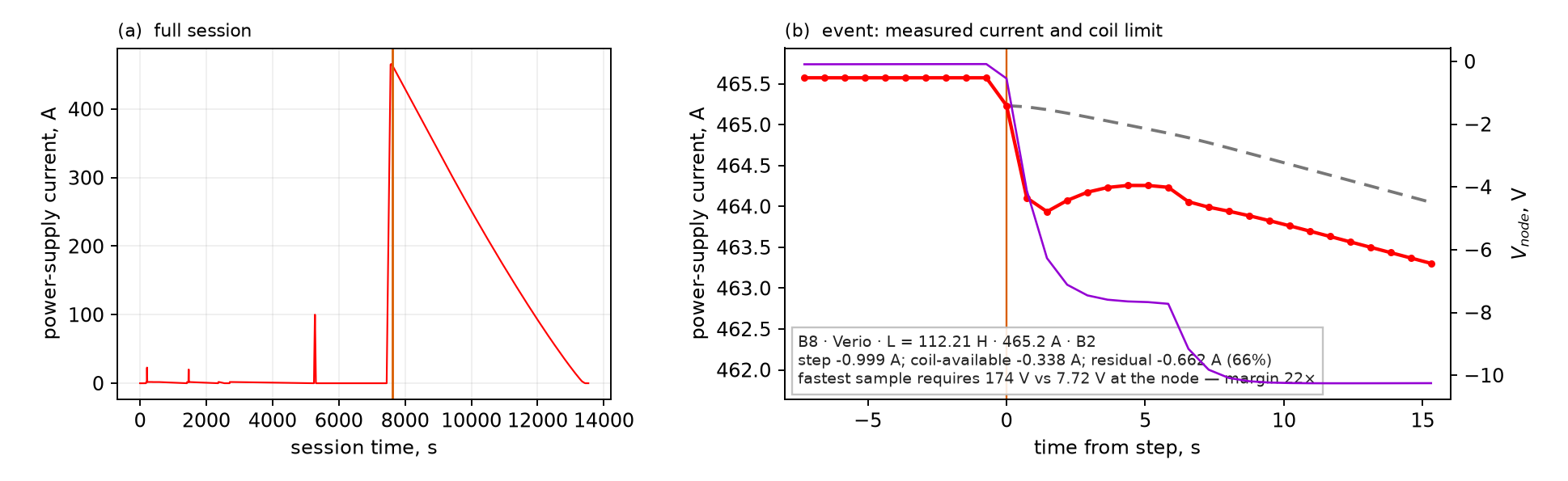}
\caption{Magnet M12 (Verio), 465.2 A, class B2.}
\label{fig:B8}
\end{figure}
\begin{figure}[!htbp]
\centering
\includegraphics[width=\linewidth]{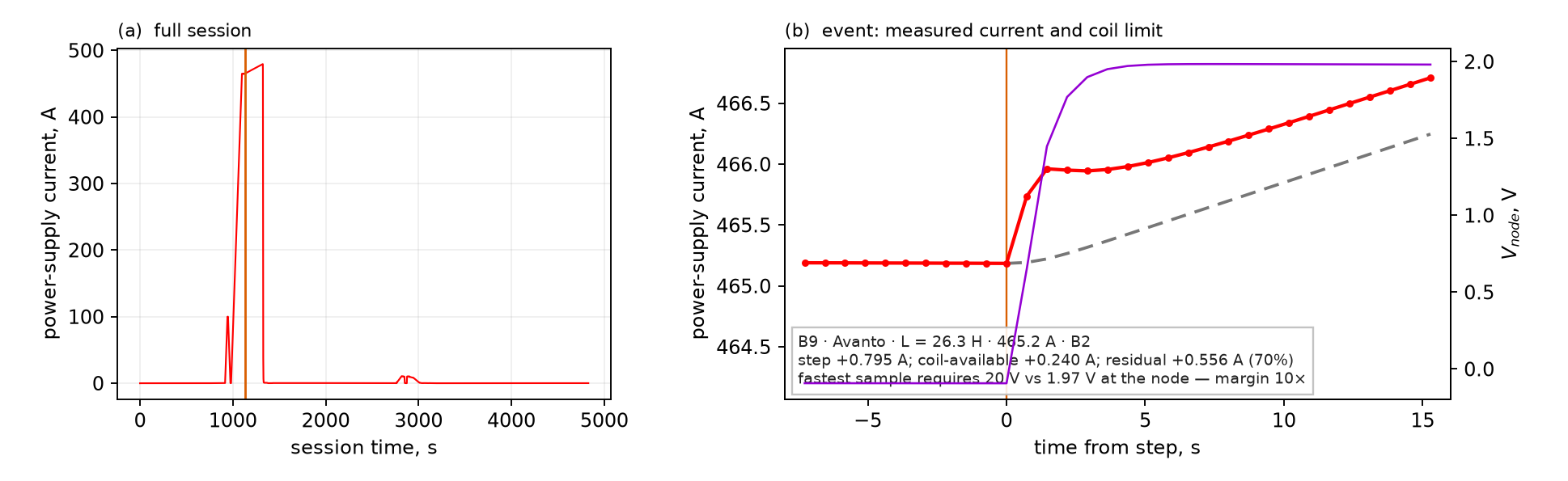}
\caption{Magnet M02 (Avanto), 465.2 A, class B2; same event as Fig. 1d.}
\label{fig:B9}
\end{figure}
\begin{figure}[!htbp]
\centering
\includegraphics[width=\linewidth]{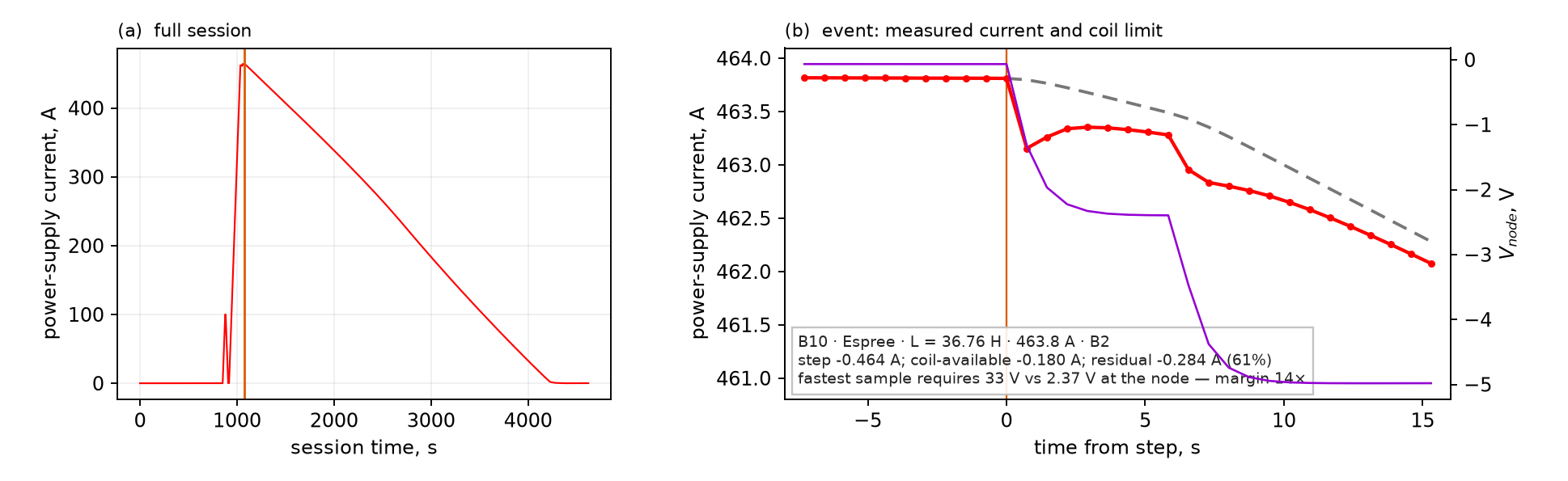}
\caption{Magnet M07 (Espree), 463.8 A, class B2.}
\label{fig:B10}
\end{figure}
\begin{figure}[!htbp]
\centering
\includegraphics[width=\linewidth]{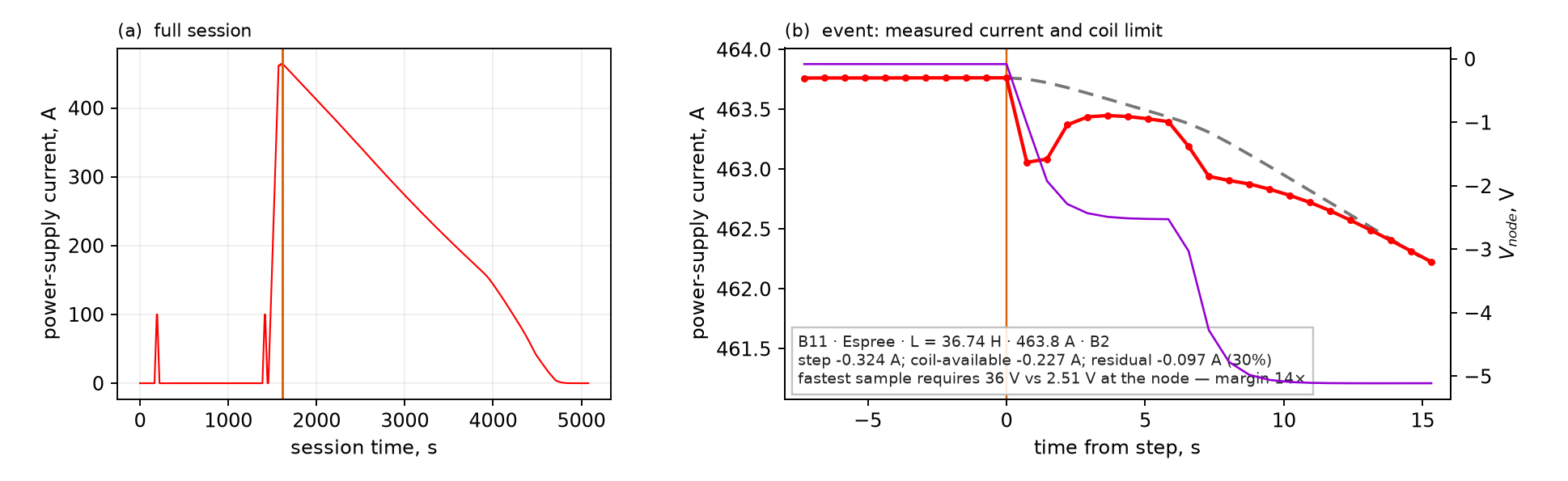}
\caption{Magnet M06 (Espree), 463.8 A, class B2.}
\label{fig:B11}
\end{figure}
\begin{figure}[!htbp]
\centering
\includegraphics[width=\linewidth]{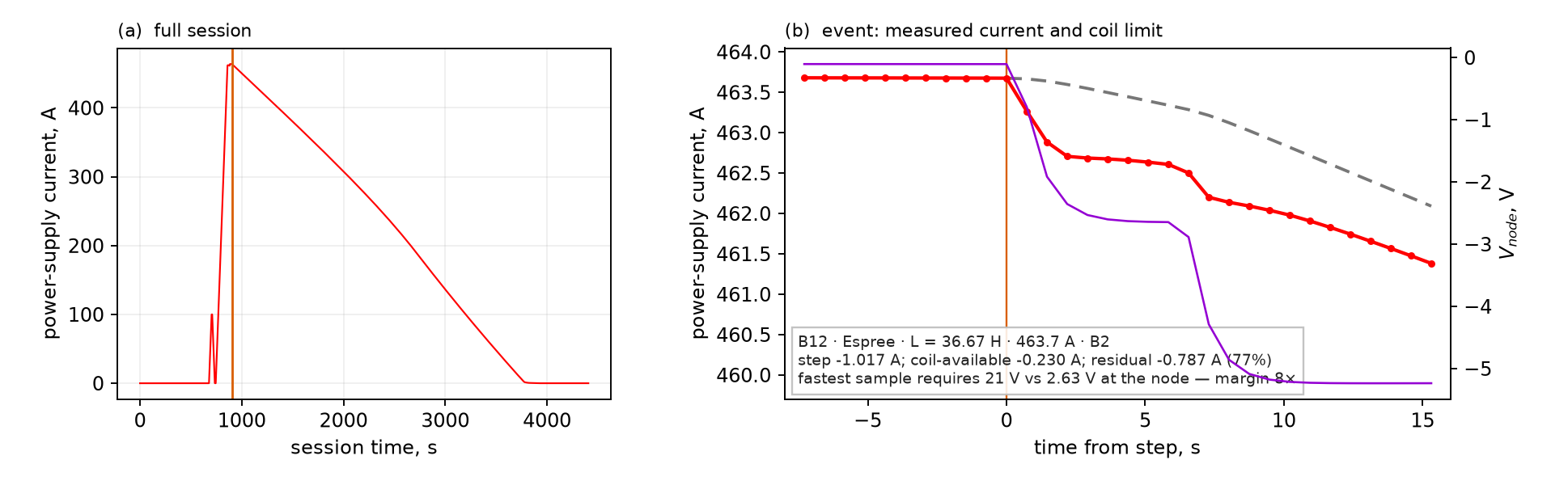}
\caption{Magnet M09 (Espree), 463.7 A, class B2.}
\label{fig:B12}
\end{figure}
\begin{figure}[!htbp]
\centering
\includegraphics[width=\linewidth]{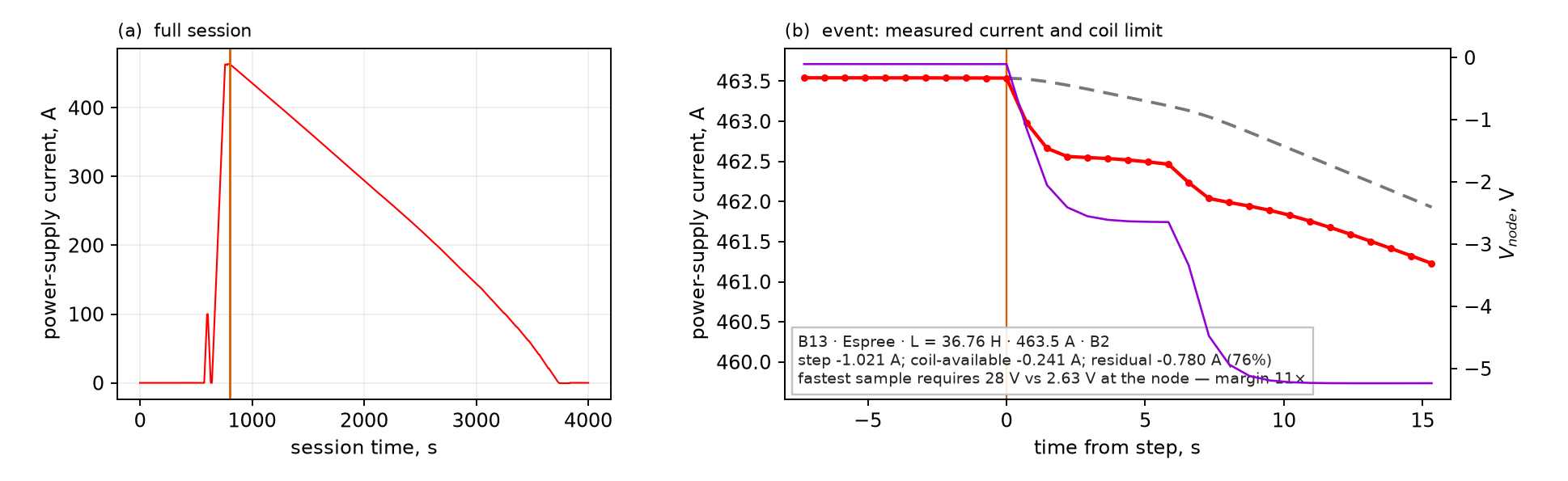}
\caption{Magnet M08 (Espree), 463.5 A, class B2.}
\label{fig:B13}
\end{figure}
\begin{figure}[!htbp]
\centering
\includegraphics[width=\linewidth]{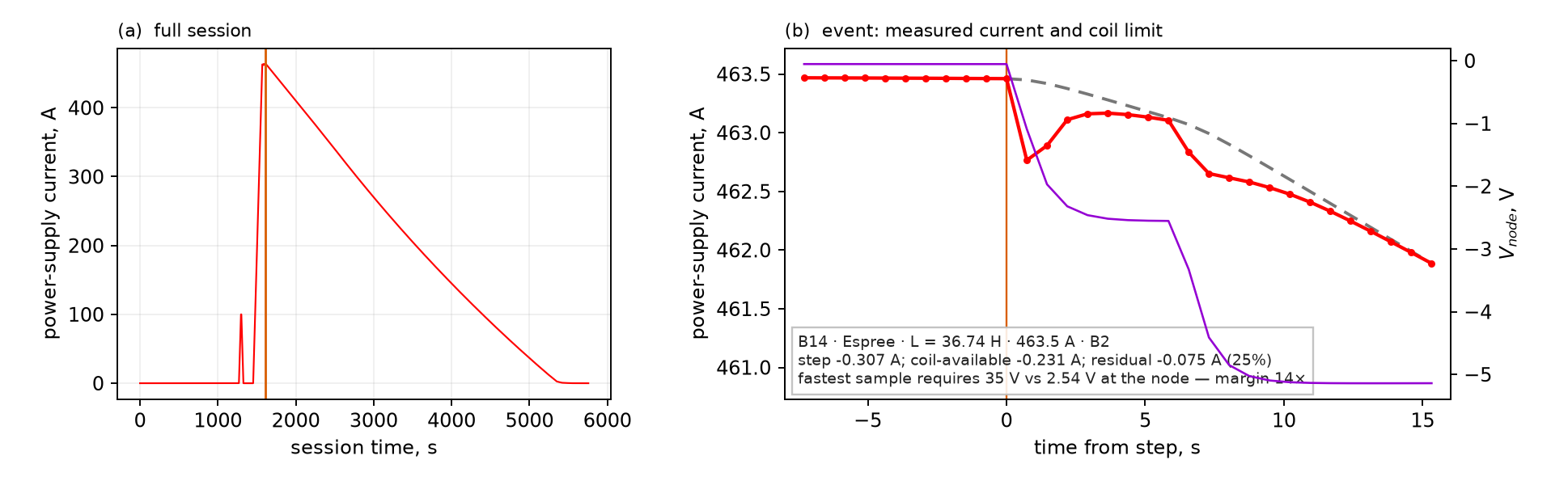}
\caption{Magnet M06 (Espree), 463.5 A, class B2; smallest residual in the corpus, 0.075 A.}
\label{fig:B14}
\end{figure}
\begin{figure}[!htbp]
\centering
\includegraphics[width=\linewidth]{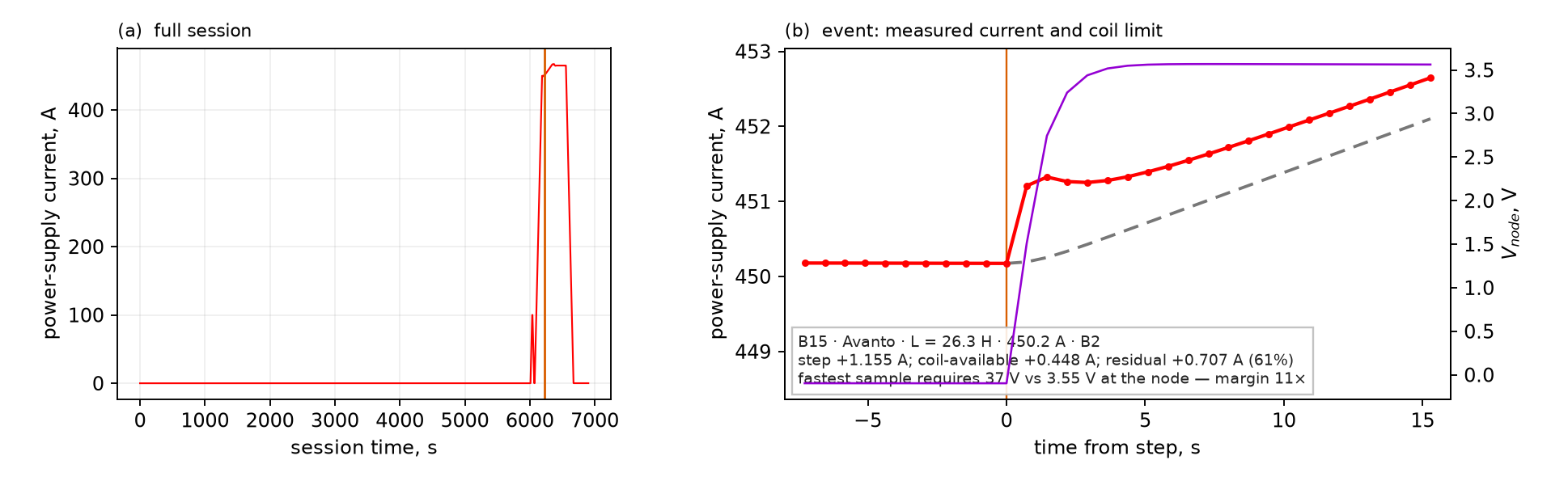}
\caption{Magnet M02 (Avanto), 450.2 A, class B2.}
\label{fig:B15}
\end{figure}
\begin{figure}[!htbp]
\centering
\includegraphics[width=\linewidth]{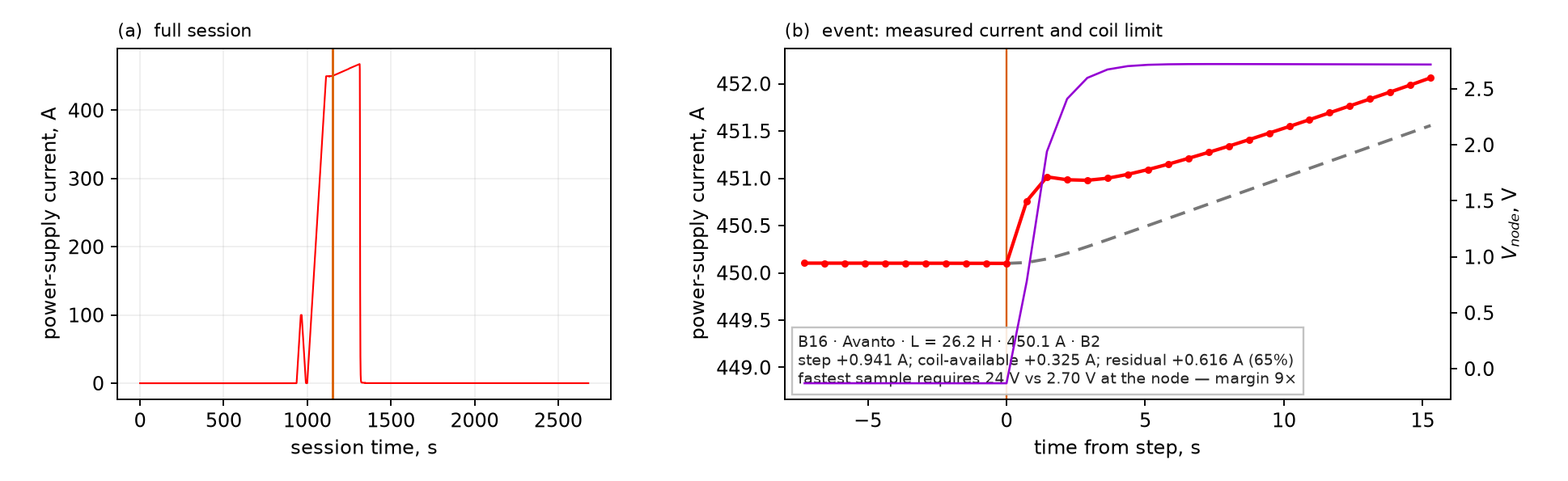}
\caption{Magnet M03 (Avanto), 450.1 A, class B2.}
\label{fig:B16}
\end{figure}
\begin{figure}[!htbp]
\centering
\includegraphics[width=\linewidth]{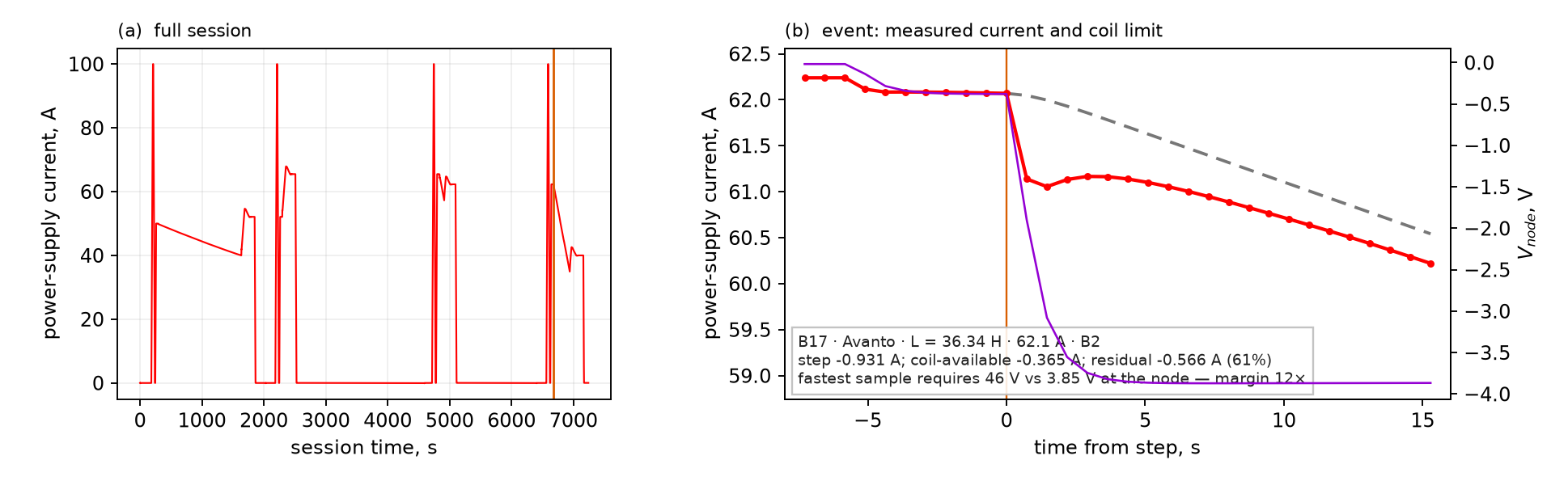}
\caption{Magnet M04 (Avanto), 62.1 A, class B2.}
\label{fig:B17}
\end{figure}
\begin{figure}[!htbp]
\centering
\includegraphics[width=\linewidth]{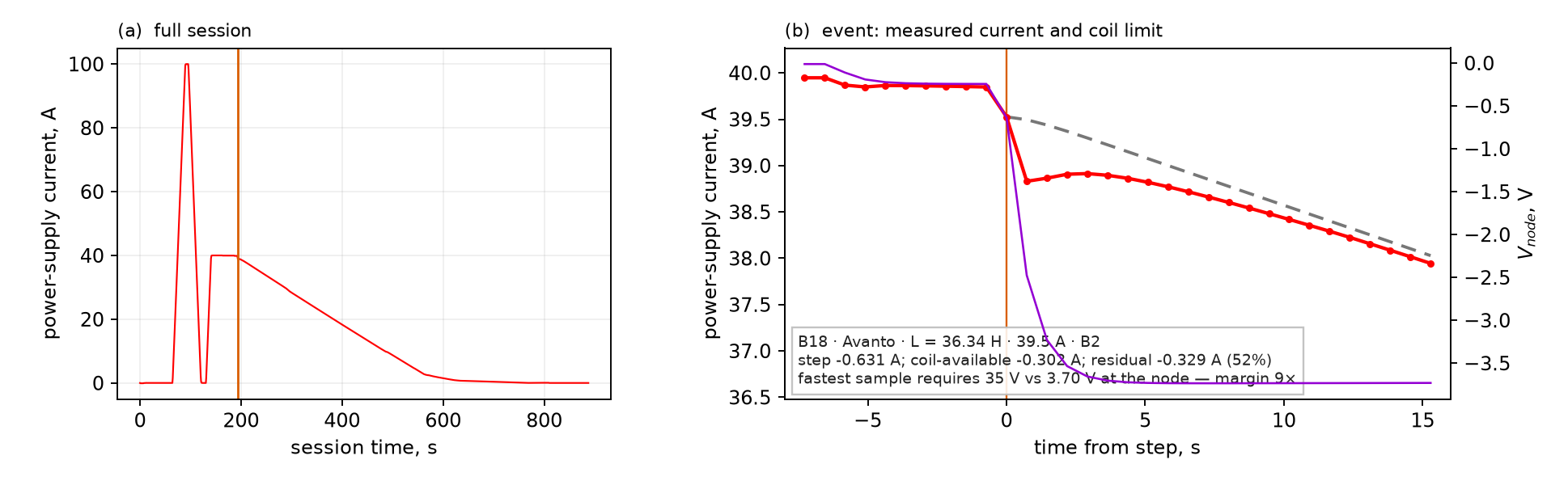}
\caption{Magnet M04 (Avanto), 39.5 A, class B2.}
\label{fig:B18}
\end{figure}
\begin{figure}[!htbp]
\centering
\includegraphics[width=\linewidth]{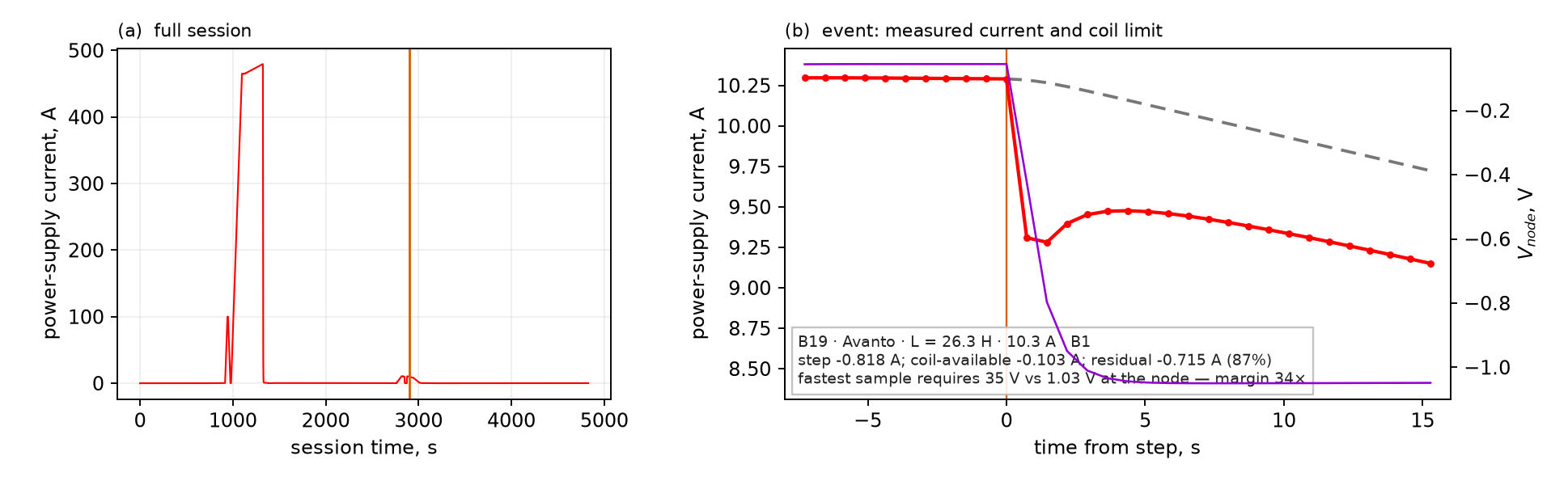}
\caption{Magnet M02 (Avanto), 10.3 A, class B1 --- the only event in which only the voltage setpoint changes; same event as Fig. 1c and Fig. 15a.}
\label{fig:B19}
\end{figure}
\FloatBarrier
\section*{Appendix C. Examples of Excluded Events}
\addcontentsline{toc}{section}{Appendix C. Examples of Excluded Events}
\setcounter{figure}{0}
\renewcommand{\thefigure}{C\arabic{figure}}
The selection in Section 6.3 removes physically ambiguous fast current changes before any balance analysis. One example of each exclusion class is shown below. In every case, a fast change in power-supply current is present, but its cause cannot be reduced to current redistribution between the coil and a parallel branch at a fixed system state; such events are therefore excluded from the corpus. Step decomposition is intentionally not calculated for them.

\begin{figure}[!htbp]
\centering
\includegraphics[width=\linewidth]{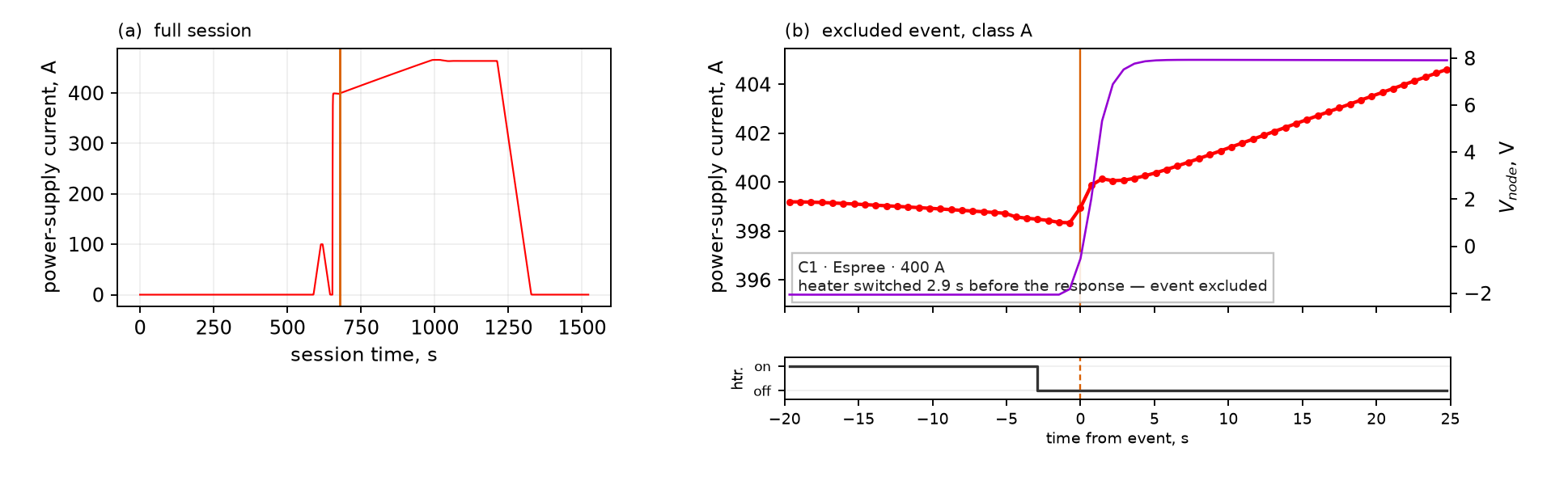}
\caption{Class A --- switching of the PCS itself: magnet M15 (Espree), 399 A. The heater switched 2.9 s before the response (lower trace: recorded heater state), so the current change cannot be separated from the change in switch state.}
\label{fig:C1}
\end{figure}
\FloatBarrier
\begin{figure}[!htbp]
\centering
\includegraphics[width=\linewidth]{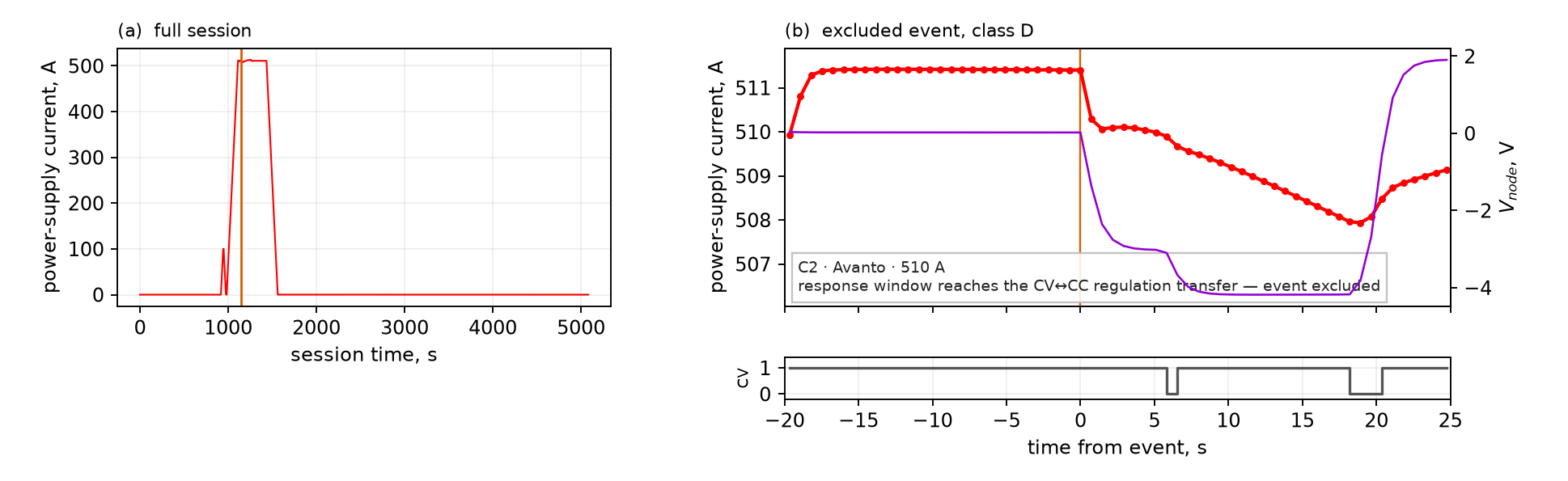}
\caption{Class D --- regulation transfer: magnet M01 (Avanto), 511 A. The response window extends into a CV$\leftrightarrow$CC transition (lower trace: voltage-regulation-mode flag), and the step is formed by the power-supply regulator. This is the twentieth candidate from the initial review that was excluded during the independent re-analysis (Appendix A).}
\label{fig:C2}
\end{figure}
\FloatBarrier
\begin{figure}[!htbp]
\centering
\includegraphics[width=\linewidth]{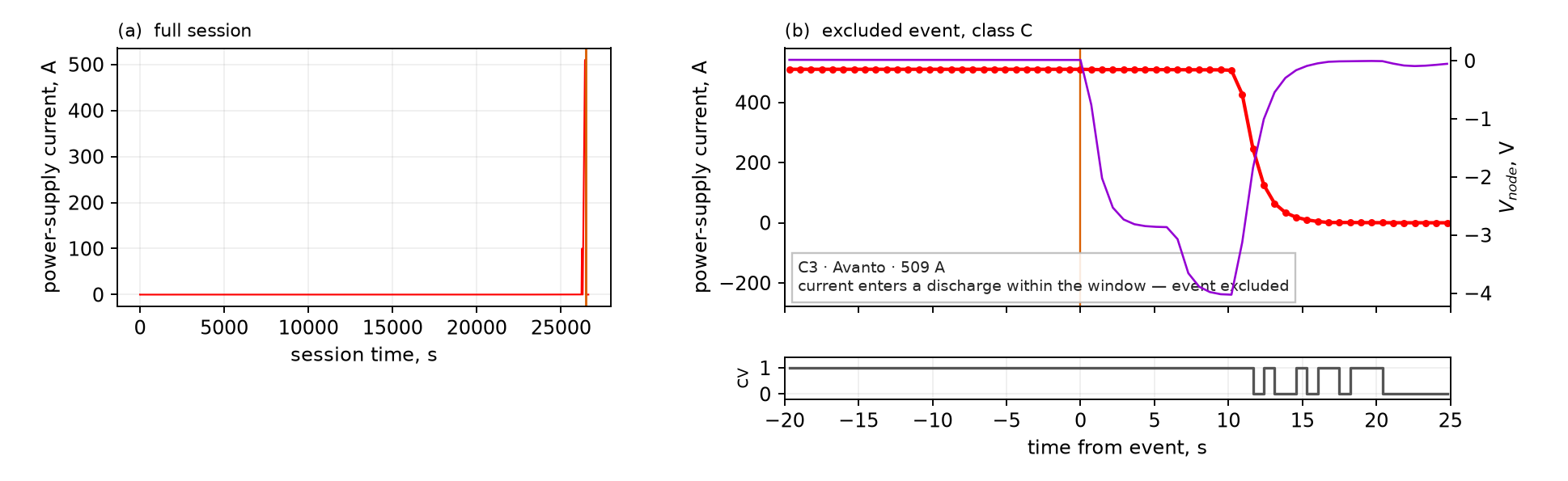}
\caption{Class C --- transition into discharge: magnet M14 (Avanto), 510 A. Within the event window, the current enters a discharge; the event is not a step on a settled operating segment.}
\label{fig:C3}
\end{figure}
\FloatBarrier
\textbf{Data availability.} Derived artifacts are available upon reasonable request; raw service archives are not distributed. Customer-identifying information and installation locations are neither used nor disclosed.

\textbf{Conflict of interest.} See the disclosure at the beginning of the manuscript.

\textbf{Author contribution.} Maneuver design, data recording, analysis, and writing: the author.

\end{document}